\documentclass{article}  
\newcommand {\bea}{\vspace{-.00in}\begin{eqnarray}}   %put -.5in for 12 pt
\newcommand {\eea}{\vspace{-.00in}\end{eqnarray}}
\newcommand {\beaa}{\vspace{-.00in}\begin{eqnarray*}}   %put -.5in for 12 pt
\newcommand {\eeaa}{\vspace{-.00in}\end{eqnarray*}}
\newcommand {\be}{\vspace{-.00in}\begin{equation}}   %put ???in for 12 pt
\newcommand {\ee}{\vspace{-.00in}\end{equation}}
\newcommand {\eps} {\epsilon}

\newcommand {\bm}  {\boldmath}
\newcommand {\del} {\delta}
\newcommand {\sig} {\sigma}
\newcommand  {\lam} {\lambda}
\newcommand  {\pr} {\prime}
\newcommand  {\kap} {\kappa}
\newcommand  {\Lam} {\Lambda}
\newcommand  {\gam} {\gamma}

\newcommand  {\veps} {\varepsilon}
\newcommand   {\pp}   {({\bf p}-{{\bf p}^\pr})^2}
\newcommand   {\easy}  {\del M_{_{1}}^2}
\newcommand    {\hard} {\del M_{_{2}}^2}
\newcommand    {\p} {\Omega_p}
\newcommand    {\ppr} {\Omega_{p^\prime}}

\newcommand    {\pig} {\left(\int_0^\pi \frac{d\omega \sin^2\omega}
{1+\cos\omega} f_{2,1}(\omega) f_{n,1}(\omega)\right)^2}
 \newcommand    {\tp} {{\tilde{\phi}}}
 
\begin{document}
%%%%%%%%%%%%%%%%%%%%%%%%%%%%%%%%%%%%%%%%%%%%%%%%%%%%%%%%%%%%%%%%%%%%%%%%%%%
%%%%%
%%%%%%%
%%%%%%%         Here's the title/abstract
%%%%%%%%%%%%%%%%%%%%%%%%%%%%%%%%%%%%%%%%%%%%%%%%%%%%%%%%%%%%%%%%%%%%%%%%%%%%%
\centerline{\large {\bf   Analytic Treatment of Positronium Spin Splittings
in Light-Front QED
}}
\vskip.1in
\centerline{Billy D. Jones\footnote{email: bjones@mps.ohio-state.edu} and Robert J. Perry}
\centerline{  Department of Physics, The Ohio State University, Columbus, Ohio 43210-1106}
\vskip.1in
\centerline{Stanis{\l}aw 
D. G{\l}azek}
\centerline{ Institute of Theoretical Physics,
Warsaw University, ul. Ho\.{z}a 69, 00-861 Warsaw, Poland}
\vskip.1in
 We study the QED bound-state problem  in
 a  light-front
  hamiltonian approach. 
 Starting with a bare cutoff QED Hamiltonian, $H_{_{B}}$, with matrix elements 
 between free states of
 drastically different energies removed, we perform a similarity
  transformation that removes
 the matrix elements between free states 
 with energy differences between the bare cutoff, $\Lambda$, and  effective
  cutoff, $\lam$ ($\lam < \Lam$). 
 This generates effective interactions in the renormalized
  Hamiltonian, $H_{_{R}}$. These effective interactions are derived to 
order $\alpha$
  in this work, with $\alpha \ll 1$. $H_{_{R}}$ is renormalized 
  by requiring it to satisfy coupling coherence.
  A nonrelativistic limit of the theory is taken, 
and the resulting Hamiltonian
 is studied  using bound-state perturbation theory (BSPT).
 The effective cutoff, $\lam^2$, is fixed, and the limit, 
 $0 \longleftarrow m^2 \alpha^2\ll
  \lam^2 \ll m^2 \alpha \longrightarrow \infty$,  is taken.
 This upper bound on $\lam^2$ places the  effects of 
 low-energy (energy transfer below $\lam$)
 emission 
 in the effective interactions in the $| e {\overline e} \rangle $ sector.
 This lower bound on $\lam^2$ insures that the nonperturbative
 scale of interest is not removed by the similarity transformation.
 As an explicit example of the general formalism introduced, we show that 
 the Hamiltonian renormalized to ${\cal O}(\alpha)$ reproduces
  the exact spectrum of spin splittings,
    with degeneracies dictated by rotational symmetry,
    for the ground state
 through  ${\cal O}(\alpha^4)$.
 The entire 
calculation is performed analytically, and gives the well known
  singlet-triplet ground state spin splitting
of positronium, $\frac{7}{6} \alpha^2 Ryd$.
We discuss remaining corrections other than the spin splittings
 and how they can be treated in calculating the spectrum with higher precision.
\vskip.2in
\noindent
PACS number(s): 11.10.Ef, 11.10.Gh, 12.20.Ds
%%%%%%%%%%%%%%%%%%%%%%%%%%%%%%%%%%%%%%%%%%%%%%%%%%%%%%%%%%%%%%%%%%%%%%%%%%%%%
%%%%%%%%%%%%
%%%%%%%%%%%%
%%%%%%%%%%%%                here's the intro!
%%%%%%%%%%%%
%%%%%%%%%%%%%%%%%%%%%%%%%%%%%%%%%%%%%%%%%%%%%%%%%%%%%%%%%%%%%%%%%%%%%%%%%%%%%%
\newpage
\centerline{{\bf I. INTRODUCTION}}
\vskip.2in

 There is much effort being put into solving for the hadronic spectrum from first principles of
$QCD$ in ($3+1$)-dimensions 
using a light-front similarity hamiltonian approach [1-5]. However, 
low-energy $QCD$ is challenging,
and a realistic analytical calculation may be impossible.
   There is a need for  exact analytical calculations that test and illustrate the
approach. 
This paper is one such calculation.
The calculation in this paper was discussed in Ref.~\cite{perrybrazil}, where the
leading order calculation was completed.
  We start from the canonical $QED_{_{3+1}}$ Hamiltonian, and  
  set up the general formalism for
 deriving the  renormalized  Hamiltonian, $H_{_{R}}$, 
 to some prescribed order in $\alpha$ (with $\alpha \ll 1$), and 
 then using BSPT to compute the low energy observables of interest to some prescribed order
 in $\alpha$.
 
 The general formulae are then applied explicitly to the spin splittings in positronium.
 Renormalization is  carried out  to order $\alpha$, and a 
 nonrelativistic limit of the theory is taken. The limit,
 \bea
  &&0 \longleftarrow m^2 \alpha^2\ll
  \lam^2 \ll m^2 \alpha \longrightarrow \infty~,
  \label{eq:limit}
  \eea
   is taken, and with 
  BSPT, a  calculation of the singlet-triplet ground state
 spin splitting of positronium to order $\alpha^4$ is carried out.
 The entire
calculation is performed
  analytically, and the well known result, $\frac{7}{6} \alpha^2 Ryd$, is obtained.  
  Our results apply to all physical positronium states with arbitrary momentum ${\cal P}$.
 
 Kaluza and Pirner have calculated the 
 singlet-triplet ground state spin splitting of positronium (neglecting annihilation channel contributions)
 to order $\alpha^4$
 in a light-front hamiltonian approach~\cite{kaluzapirner},
 and they 
 obtained correct results numerically, but were forced to make ad hoc assumptions because
 their Hamiltonian depended on the full eigenvalue of the problem.
 We avoid these assumptions 
 in our approach,
 and perform the calculation analytically.
 %%%%%%%%%%%%%%%%%%%%%%%%%%%%%%%%%%%%%%%%%%%%%%%%%%%%%%%%%%%%%%%%%%%%%%%%%%%%%
%%%%%%%%%%%%
%%%%%%%%%%%%
%%%%%%%%%%%%                here's "the similarity hamiltonian approach"
%%%%%%%%%%%%
%%%%%%%%%%%%%%%%%%%%%%%%%%%%%%%%%%%%%%%%%%%%%%%%%%%%%%%%%%%%%%%%%%%%%%%%%%%%%%
\vskip.2in
\noindent
\centerline{{\bf II. THE SIMILARITY HAMILTONIAN APPROACH}}
\vskip.2in

The starting point in the similarity hamiltonian approach
  is a 
bare cutoff continuum Hamiltonian of physical interest, $H_{_{B}}$, with 
energy widths\footnote{The ``energy
difference" between the free states in a matrix element of a Hamiltonian is defined to be its ``energy 
width."} restricted to be below the bare cutoff, $\Lam$. A similarity  transformation 
(unitary here) 
is defined that acts on $H_{_{B}}$ and restricts the 
energy widths in the final  Hamiltonian, $H_{_{R}}$, to be below the effective
 cutoff, $\lam$.\footnote{In the initial setup 
of the similarity transformation, $\lam$ and $\Lam$ will be used as a shorthand
for $\frac{\lam^2}{{\cal P}^+}$ and $\frac{\Lam^2}{{\cal P}^+}$ respectively,
where ${\cal P}^+$ is the total longitudinal momentum of the physical state of interest,
and $\lam^2$ and $\Lam^2$ have dimension $({\rm mass})^2$.}
All the energy changes between $\Lam$ and $\lam$  are ``integrated out" and replaced by
effective interactions in $H_{_{R}}$. 
The initial Hamiltonian must then be adjusted so that $H_{_{R}}$ satisfies coupling 
coherence~\cite{cc1}, 
which produces a renormalized Hamiltonian order by order
  in the running couplings, $e_\lam$ and $m_\lam$.
 Coupling coherence can be realized by requiring
  a Hamiltonian at one scale to equal a Hamiltonian 
 at a new scale after changing the explicit scale dependence 
 in the Hamiltonian and the implicit
 scale dependence in a finite number of  independent running couplings. In addition, 
  all the dependent  couplings are required to 
  vanish when the  independent  marginal couplings 
  are taken to zero. 
 
 The second step of this similarity hamiltonian approach is the diagonalization
 of $H_{_{R}}$. First, for this QED calculation,  a nonrelativistic limit of  $H_{_{R}}$
 is taken. 
 This is reasonable because the bound-state electron momenta
  (equivalent formulae hold for the positron) satisfy:
 \bea
 \frac{p^+_{electron}}{{\cal P}^+} &=& \frac{1}{2} + {\cal O}(\alpha)\;,\label{eq:nr1}\\
 \frac{p^\perp_{electron}}{m} &=& {\cal O}(\alpha)\;,\label{eq:nr2}\\
 \alpha &\ll& 1\;,
 \eea
 where $m$ is the renormalized
 electron mass and ${\cal P}^+$ is the total longitudinal momentum of positronium.\footnote{The
  reader unaccustomed to light-front coordinates should consult Appendix~A.}
 
   Next, the Hamiltonian is divided into a nonperturbative and
   perturbative part,
  \bea
  H_{_{R}}&=& {\cal H}_o+(H_{_{R}}-{\cal H}_o) \equiv {\cal H}_o+{\cal V}\;.
  \eea
  This is a standard trick for hamiltonian problems, and will  
  work best if the lowest order spectrum of $H_{_{R}}$ is well approximated 
  by the spectrum of ${\cal H}_o$.  Phenomenological input can be used 
  to determine ${\cal H}_o$. The main point is that
  if the spectrum of ${\cal H}_o$ differs too much from the lowest order spectrum of $H_{_{R}}$, 
   the subsequent BSPT will 
  not converge  rapidly (if at all).
  
  We proceed to estimate a lower bound on $\lam$ where the nonperturbative bound state effects
of $H_\lam$ enter. This lower bound naturally appears at the physical binding energy of interest:
\bea
\frac{\lam^2}{{\cal P}^+}&\gg&  \frac{\left|M_N^2-(2 m)^2\right|}{{\cal P}^+}
\;,\label{eq:2}
\eea
where $M_N$ is the mass of the physical state of interest.\footnote{$M_N^2 \equiv
(2 m +B_N)^2$, where $N$ labels all the quantum numbers 
and $-B_N$ is the binding energy of the physical state of interest.}
Given this lower bound we next proceed with an estimate of an upper bound on $\lam$ that will 
allow
us to   obtain  our approximate spectrum 
entirely within the $|e {\overline e} \rangle$ sector. The fact that this upper bound arises at all
 is one of the utilities of the similarity
hamiltonian approach.
This upper bound is
\bea
\frac{\lam^2}{{\cal P}^+}&\ll&q^{-}_{_{photon}}\;,
\eea
where $q^{-}_{_{photon}}$ is the dominant energy of  emitted or absorbed photons.

 Given this range of $\lam$ that allows the mass eigenstates of the theory to be dominated by the few-body
 sector of interest, and at the same time, does not remove the nonperturbative
 bound state physics of interest,  we  proceed with the BSPT and
  calculate all relativistic corrections to the zeroth order spectrum
to some specified order in $\alpha$. The utility of this paper is that 
the relativistic corrections  to order $\alpha^4$ in the ground state spin splittings can
be calculated analytically. In the following subsections we  describe the similarity 
hamiltonian approach more explicitly.
%%%%%%%%%%%%%%%%%%%%%%%%%%%%%%%%%%%%%%%%%%%%%%%%%%%%%%%%%%%%%%%%%%%%%%%%%%%%%
%%%%%%%%%%%%
%%%%%%%%%%%%
%%%%%%%%%%%%                here's "step one: derive Hsigma"
%%%%%%%%%%%%
%%%%%%%%%%%%%%%%%%%%%%%%%%%%%%%%%%%%%%%%%%%%%%%%%%%%%%%%%%%%%%%%%%%%%%%%%%%%%%
\vskip.2in
\noindent
\centerline{{\bf A. Step one:  derivation of {\boldmath $H_{_{R}}$}}}
\vskip.2in

A self-contained discussion of the derivation of the
renormalized Hamiltonian, $H_{_{R}}$, will now be given.  The similarity
transformation we use to derive our effective interactions was developed by
G{\l}azek and Wilson
 \cite{glazekwilson1}, and separately by Wegner \cite{wegner}.
 An early application of this approach \cite{longpaper} was a weak-coupling treatment
 of QCD.

The approach starts with a  definition of the bare
 Hamiltonian:
\bea
H_{_{B}}&\equiv&h+v_{_{\Lam}} \;,\label{eq:hap1}\\
v_{_{\Lam}}&\equiv& f_{_{\Lam}} {\overline v}_{_{\Lam}}\;,\\
{\overline v}_{_{\Lam}}&\equiv& v_{can}+\del v_{_{\Lam}}\;,\\
H_{can}&\equiv&h+v_{can}
\;,
\eea
where $h$ is the free Hamiltonian, $H_{can}$ is the canonical Hamiltonian, $f_{_{\Lam}}$ is 
a regulating function, and $\del v_{_{\Lam}}$ are counterterms defined through the process
of renormalization.
The canonical Hamiltonian, $H_{can}$, is written in terms of renormalized parameters and will
be specified at the end of this section. The counterterms, $\del v_{_{\Lam}}$, are fixed
by coupling coherence. Coupling coherence will be explained further below.
 
  The free Hamiltonian, $h$, is given by:
\bea
h&\equiv& \int_p \sum_s \left \{\left(b_s^\dagger(p) b_s(p)+
 d_s^\dagger(p) d_s(p)\right)
 \left(\frac{{p^\perp}^2+m^2}{p^+}\right)+
  a_s^\dagger(p) a_s(p)
 \left(\frac{{p^\perp}^2}{p^+}\right)\right \}
 \label{eq:hap2}\;,\\
 &&~~~~~~~~~~~~~~~~~~~~~~~~~h |i\rangle= \veps_i |i\rangle\;\;,\;\;\sum_i |i\rangle\langle i | = 1 \;,
 \label{eq:11}\eea
 where the sum over $i$ implies a sum over all Fock sectors and spins, and integrations
 over all momenta in the respective free states.
 We use the shorthand $\int_p = \int \frac{d^2p^\perp dp^+ \theta(p^+)}
 {16 \pi^3 p^+}$. 
$m$ is the renormalized fermion mass.
 
 The regulating function, 
$f_{_{\Lam}}$, is defined to act in 
the following way:
\bea
\langle i | f_{_{\Lam}} {\overline v}_{_{\Lam}} |j \rangle &\equiv& f_{_{\Lam ij}} \langle i | 
{\overline v}_{_{\Lam}} |j\rangle \equiv f_{_{\Lam ij}} {\overline v}_{_{\Lam ij}}~,\\
f_{_{\Lam ij}}&\equiv& \theta(\Lam-|\Delta_{ij}|)\;\;,\;\;\Delta_{ij}\equiv \veps_i-\veps_j
\;.
\eea
Note that this choice of a step function is not necessary
and can lead to pathologies, however it is useful for doing analytical
calculations.

Next, a similarity transformation is defined that acts on $H_{_{B}}$ and restricts the energy
widths in the renormalized Hamiltonian, $H_{_{R}}$, to be below the effective  cutoff, $\lam$. 
This transformation allows  recursion relationships 
to be set up for $H_{_{R}}$, which can be written in the following general form:
\bea
H_{_{R}}&=& h+v_\lam \label{eq:hap3}~,\\
v_\lam&\equiv&f_\lam {\overline v}_\lam\;,\\
{\overline v}_\lam&=&{\overline v}_\lam^{^{(1)}}+{\overline v}_\lam^{^{(2)}}+\cdots
\;,
\eea
where the superscripts imply the respective order in $v_{can}$. 
 
 Now, starting with the above bare Hamiltonian, we will describe this procedure more explicitly.
The similarity transformation is 
defined
to act on a bare cutoff continuum Hamiltonian, $H_{_{B}}$, in the following way:
\bea
H_{_{R}}&\equiv& S(\lam ,\Lam) H_{_{B}} S^\dagger(\lam, \Lam)\;,\label{eq:star}
\\
S(\lam, \Lam)  S^\dagger(\lam, \Lam)&\equiv&S^\dagger(\lam, \Lam) S(\lam, \Lam)  \equiv 1
\;.
\eea
 This transformation is unitary, so  $H_{_{B}}$ and $H_{_{R}}$ have the
 same spectrum:
 \bea
 H_{_{B}} |\Psi_{_{B}}\rangle&=& E |\Psi_{_{B}}\rangle\;,\\
 \underbrace{S(\lam ,\Lam) H_{_{B}} S^\dagger(\lam ,\Lam)}_{H_{_{R}}}
 \underbrace{ S(\lam, \Lam)|\Psi_{_{B}}\rangle}_{|\Psi_{_{R}}\rangle}&=&E \underbrace{S(\lam, \Lam) 
 |\Psi_{_{B}}\rangle}_{|\Psi_{_{R}}\rangle}
 \;.
 \eea
 Therefore, $E$ is independent of
 the effective cutoff,
  $\lam$, if an exact transformation is made.
 $E$ is also independent of the bare
 cutoff, $\Lam$, after the Hamiltonian is renormalized. 
 
 To put the equations in a differential framework, note that
  Eq.~(\ref{eq:star}) is equivalent to the following equation
 \bea
 \frac{d H_{_{R}}}{d \lam}&=& \left [ H_{_{R}},T_\lam\right] \label{eq:star2}\;,
\label{eq:wnbv} \eea
 with 
 \bea
 S(\lam ,\Lam)&\equiv&{\cal T} \exp \left(\int_\lam^\Lam T_{\lam^\pr} d \lam^\pr\right)\;,
 \eea
 where `${\cal T}$' orders operators from left to right in order of increasing energy scale, 
 $\lam^\pr$.
 Eq.~(\ref{eq:wnbv}) is a first order differential equation, thus one boundary condition
 must be specified to obtain its solution. This boundary condition is the bare Hamiltonian:
  $H_{_{R}} |_{_{\lam \rightarrow \Lam}} \equiv H_{_{B}}$.  $H_{_{B}}$ is determined by
  coupling coherence.
  Now we must specify $T_\lam$, the anti-hermitian ($T_\lam^\dagger=-T_\lam$) generator of
 energy width transformations. To define $T_\lam$ note that it is enough to specify how 
 ${\overline v}_\lam$ and $h$ change with the energy scale $\lam$. This is seen
  by writing out Eq.~(\ref{eq:star2}) more explicitly using Eq.~(\ref{eq:hap3}):
  \bea
\frac{d h}{d\lam}+\frac{d}{d\lam}\left(f_\lam {\overline v}_\lam \right)&=&
[h,T_\lam]+[v_\lam,T_\lam]\;.\label{eq:19}
\eea
We solve this perturbatively in $v_{can}$, choosing the transformation
so that $h$ is independent of $\lam$.  Also, we demand that
  $T_\lam$ and $v_\lam$ do not contain any small energy denominators.  Thus we define:
\bea
\frac{d h}{d \lam}&\equiv&0\;,\label{eq:star4}\\
\frac{d {\overline v}_\lam}{d \lam}&\equiv& [v_\lam,T_\lam]\;.
\label{eq:star3}
\eea
Eq.~(\ref{eq:star3}) is a choice such that   
$T_\lam$ and consequently $v_\lam$
do not allow any small energy denominators.
These additional constraints determine $T_\lam$ and ${\overline v}_\lam$, which  are 
 given by the following 
equations:
\bea
 \left[h,T_\lam\right]
&=&
 \overline{v}_\lam \frac{d f_\lam}{d \lam} -
 {\overline f}_\lam [v_\lam,T_\lam]\;,  \label{eq:22}\\
{\overline v}_\lam &=& v_{can}+\del v_{_{\Lam}} -\int_\lam^\Lam [ v_{\lam^\pr},T_{\lam^\pr}]
d \lam^\pr
\;,\label{eq:23}
\eea
 where $f_\lam+{\overline f}_\lam\equiv 1
 \;$\@.
 Eqs.~(\ref{eq:22}) and (\ref{eq:23}) follow from Eqs.~(\ref{eq:19})-(\ref{eq:star3})
 and the boundary condition $H_{_{R}} |_{_{\lam \rightarrow \Lam}} \equiv H_{_{B}}$.
 Now we solve Eqs.~(\ref{eq:22}) and (\ref{eq:23}) for $T_\lam$ and $\overline{v}_\lam$.
 Given Eq.~(\ref{eq:hap3}), we need to determine ${\overline v}_\lam$,
and $H_{_{R}}$ is known. The solution to Eqs.~(\ref{eq:22}) and (\ref{eq:23}) is:
\bea
&&{\overline v}_{\lam}=
{\overline v}_{\lam}^{^{(1)}}
+{\overline v}_{\lam}^{^{(2)}}+\cdots
\label{eq:yes}~,
\\
&&T_{\lam}=
T_{\lam}^{^{(1)}}
+T_{\lam}^{^{(2)}}+\cdots~,\\
&&\del v_{_{\Lam}} = \del v_{_{\Lam}}^{^{(2)}}
+\del v_{_{\Lam}}^{^{(3)}} +\cdots\;,
\eea
where the superscripts imply the respective order in the canonical interaction, $v_{can}$, and
these quantities are given by:
\bea
 {\overline v}_\lam^{^{(1)}}&=& v_{can}
~, \\
 \left [ h , T_\lam^{^{(1)}} \right]&=& v_{can} 
 \frac{d f_\lam}{d \lam}
 ~,\\
 \overline{v}_{\lam}^{^{(2)}}&=& 
 -
\int_{\lam}^{\Lam}
d \lam^\prime [ v_{\lam^\prime}^{^{(1)}} , T_{\lam^\prime}^{^{(1)}}]
+ \delta v_{_{\Lam}}^{^{(2)}}
 ~,\label{eq:vtwo}\\
 \left[h,T_\lam^{^{(2)}}\right]
&=&
 \overline{v}_\lam^{^{(2)}} \frac{d f_\lam}{d \lam} -
 {\overline f}_\lam [v_\lam^{^{(1)}},T_\lam^{^{(1)}}] 
 ~,\\
 \overline{v}_{\lam}^{^{(3)}}&=& 
 -
\int_{\lam}^{\Lam}
d \lam^\prime \left(\left[ v_{\lam^\prime}^{^{(1)}} 
, T_{\lam^\prime}^{^{(2)}}\right] +
\left[ v_{\lam^\prime}^{^{(2)}} 
, T_{\lam^\prime}^{^{(1)}}\right]\right)
+ \delta v_{_{\Lam}}^{^{(3)}} 
~,\\
\left[h,T_\lam^{^{(3)}}\right]
&=&
 \overline{v}_\lam^{^{(3)}} \frac{d f_\lam}{d \lam} -
 {\overline f}_\lam \left(\left[ v_{\lam}^{^{(1)}} 
, T_{\lam}^{^{(2)}}\right] +
\left[ v_{\lam}^{^{(2)}} 
, T_{\lam}^{^{(1)}}\right]\right) 
~,\\
&\vdots&\;.\nonumber
 \eea
 A general form of these effective interactions is:
  \bea
 {\overline v}_\lam^{^{(i)}}&=&-\sum_{j,k=1}^\infty \del_{(j+k,i)}
 \int_\lam^\Lam d\lam^\pr[ v_{\lam^\prime}^{^{(j)}} , T_{\lam^\prime}^{^{(k)}}]
+ \delta v_{_{\Lam}}^{^{(i)}}
 \;,
 \eea
 for $i=2,3, \cdots$, with
  ${\overline v}_\lam^{^{(1)}}=v_{can}$.
 
 $H_{_{R}}$ is renormalized by requiring it 
 to satisfy  coupling coherence ~\cite{cc1}.
A coupling coherent Hamiltonian satisfies:
\bea
 S(\lam ,\Lam) H_{_{B}}(\Lam, e_\Lam, m_\Lam,
c(e_\Lam, m_\Lam)) S^\dagger(\lam ,\Lam)
&=& H_{_{B}}(\lam, e_\lam, m_\lam,c(e_\lam, m_\lam))\;,
\label{eq:cc}
\eea
with the additional requirement that all dependent couplings (only one is shown in the argument of the 
Hamiltonians for simplicity) vanish when the independent marginal
couplings are taken to zero. Note that 
there are only  a finite number of independent couplings.  
 This  is a highly non-trivial constraint on the
theory and to date has only been solved perturbatively. In this paper, Eq.~(\ref{eq:cc}) is solved
to order $e^2$, which turns out to be fairly simple because $e$ does not run until order $e^3$.

Now we write  the solution to second order
 for ${\overline v}_\lam^{^{(2)}}$.  From Eq.~(\ref{eq:vtwo}) we obtain:
\bea
  {\overline v}_{\lam ij}^{^{(2)}}&=&
  \sum_{k} {\left(v_{can}\right)}_{ik} 
  {\left(v_{can}\right)}_{kj} \left(
  \frac{g^{(\lam \Lam)}_{ikj}}{\Delta_{ik}} +
  \frac{g^{(\lam \Lam)}_{jki}}{\Delta_{jk}}
  \right) +\del v_{\Lam ij}^{^{(2)}}\;,\\
  {\rm where}&&g^{(\lam \Lam)}_{ikj} \equiv
  \int_\lam^\Lam d \lam^\prime f_{\lam^\prime jk} 
  \frac{d f_{\lam^\prime ki}}{d \lam^\prime}\;.
  \label{eq:g}
  \eea
  $\del v_{\Lam ij}^{^{(2)}}$ will be determined (in \S III) by requiring 
  the conditions of coupling coherence to be satisfied.
  These previous equations are valid for an arbitrary similarity
  function, $f_\lam$. In this work we will use
   $f_{\lam i j} = \theta(\lam-|\Delta_{ij}|)$ (a
   convenient choice for doing analytical calculations). This gives:
  \bea
  g^{(\lam \Lam)}_{ikj}&=&
  \left( f_{\Lam ik}-f_{\lam ik}
  \right) \Theta_{ikj}\;,\\
  \Theta_{ikj} &\equiv&
  \theta \left( |\Delta_{ik}|-|\Delta_{kj}|
  \right)\;.
  \label{eq:bigtheta}
  \eea
 
  To complete this section we write the canonical QED Hamiltonian.
 We
start by dividing $H_{_{can}}$ into free and interacting parts:
\bea
H_{can}&\equiv&h+v_{can}\;,
\eea
where $h$ is given by Eq.~(\ref{eq:hap2}). Starting with the QED lagrangian
($e > 0$, i.e. the charge of the electron is $-e$):
\bea
&&~~~~~~~~~~{\cal L}_{_{QED}}= -\frac{1}{4}F_{\mu \nu} F^{\mu \nu} +
\overline{\psi}(i \not\!\! {D } - 
 m )\psi\;,\\
&&{\rm with}~~~~~~~~~~\; i \not\!\!{D} \psi\;=\;\gamma^\mu (i
\partial_{\mu}+ e A_{\mu})
\psi\;,\nonumber
\eea
in a fixed gauge, $A^{+}=0$,  the
constrained degrees of freedom are removed explicitly, producing $v_{can}$.
For details of the derivation see \S IV.A of Ref.~\cite{longpaper}. We use
the two-component representation chosen by Zhang and Harindranath~\cite{threeharipapers}.
Below we write the resulting Hamiltonian completely. The 
field operator expansions and light-front conventions followed in this paper can 
be found in Appendix A.  The
canonical  Hamiltonian   is
\bea
P^{-}_{can}&\equiv&H_{can}\equiv h+v_{can}\;,
\\
v_{can}&\equiv&
 \int d^2 x^{\perp}dx^- {\cal H}_{int}\;,
\label{eq:46} 
\eea
 where
 \bea
 &&{\cal H}_{int}={\cal H}_{ee\gamma}+{\cal H}_{ee\gamma\gamma}+
{\cal H}_{eeee}\;,
\eea
and
\bea
&& {\cal H}_{ee\gamma}=e\xi^{\dag}\left\{-2(\frac{\partial^{\perp}}
{\partial^+}\cdot A^{\perp})+\sig \cdot A^{\perp} 
\frac{\sig \cdot \partial^{\perp}+m}{\partial^+}+
\frac{\sig \cdot \stackrel{\!\!\leftarrow}
{\partial^{\perp}}+m}{\stackrel{\!\!\leftarrow}{\partial^+}} 
\sig \cdot A^{\perp}
\right\}\xi
\;,\label{eq:48}\\
&&{\cal H}_{ee\gamma\gamma}=
-i e^2\left\{\xi^{\dag} \sig \cdot A^{\perp}\frac{1}{\partial^+}
 (\sig \cdot A^{\perp} \xi)\right\}
\;, \\
 &&{\cal H}_{eeee}= 2 e^2 \left\{ \left[
 \frac{1}{\partial^+} (\xi^{\dagger} \xi) \right] \left[
 \frac{1}{\partial^+} (\xi^{\dagger} \xi)\right] \right\}\;.
 \eea
 Note:  $ i\;=\;1,2\;$only; e.g.,  $\sig \cdot \partial^{\perp}=
\sig^i \partial^i = \sig^1 (-\partial_1)+\sig^2(-\partial_2)$;
$\{\sig^i\}$ are the standard $2\times2$ Pauli matrices. Also,
$h$ is given by Eq.~(\ref{eq:hap2}).
%%%%%%%%%%%%%%%%%%%%%%%%%%%%%%%%%%%%%%%%%%%%%%%%%%%%%%%%%%%%%%%%%%%%%%%%%%%%%
%%%%%%%%%%%%
%%%%%%%%%%%%
%%%%%%%%%%%%                here's "step two: solve for the spectrum
%%%%%%%%%%%%                        of  Hsigma"
%%%%%%%%%%%%
%%%%%%%%%%%%%%%%%%%%%%%%%%%%%%%%%%%%%%%%%%%%%%%%%%%%%%%%%%%%%%%%%%%%%%%%%%%%%%
\vskip.2in
\noindent
\centerline{{\bf B. Step two:  diagonalization of {\bm $H_{_{R}}$}}}
\vskip.2in

The second step in the similarity hamiltonian approach is to
solve for the spectrum of $H_{_{R}}$. 
The Schr{\"o}dinger equation for eigenstates of $H_{_{R}}$  is:
\bea
\sum_j \langle i |H_{_{R}}|j \rangle \langle j | \Psi_{_{R, N}}({\cal P})\rangle&=&
E_{_{N}} \langle i | \Psi_{_{R ,N}}({\cal P})\rangle\;.\label{eq:sad}
\eea
See Eq.~(\ref{eq:11}) and the comment immediately following it for an explanation of the notation.
`$N$' labels all the quantum numbers of the state, and is discrete for bound states and continuous
for scattering states.
 $E_{_{N}}\equiv\frac{{{\cal P}^\perp}^2+M_{_{N}}^2}{{\cal P}^+}
$, $M_{_{N}}^2\equiv(2 m+B_{_{N}})^2$, and `${\cal P}$' is the total momentum of the
 state of physical interest
(for this paper, positronium).
 
  Solving this eigenvalue 
equation exactly is not feasible, because  all
sectors are still coupled:
\bea
|\Psi_{_{R ,N}}({\cal P})\rangle&=&\sum_i |i\rangle \langle i |\Psi_{_{R ,N}}({\cal P})\rangle =
\sum_{i^\pr}|e {\overline e} ( i^\pr)\rangle \langle e 
{\overline e} (i^\pr)|\Psi_{_{R, N}}({\cal P})\rangle~+\nonumber\\
&&~~~~~+\sum_{i^\pr}|e {\overline e} \gam( i^\pr)\rangle \langle e 
{\overline e}\gam (i^\pr)|\Psi_{_{R, N}}({\cal P})\rangle+
\sum_{i^\pr}|e {\overline e} e {\overline e}( i^\pr)\rangle \langle e 
{\overline e}e {\overline e} (i^\pr)|\Psi_{_{R, N}}({\cal P})\rangle+\cdots\;.
\eea
We  divide $H_{_{R}}$ into two pieces:
\bea
H_{_{R}}&=& {\cal H}_o+\left(H_{_{R}}-{\cal H}_o\right)\equiv{\cal H}_o+{\cal V}\;,
\eea
diagonalize ${\cal H}_o$ exactly,
and calculate corrections to the spectrum of ${\cal H}_o$ in BSPT
with ${\cal V}$. The ${\cal H}_o$ we choose for positronium is:
\bea
{\cal H}_o&=&h+\sum_{i^\pr j^\pr} | e {\overline e} (i^\pr)\rangle\langle
e{\overline e}(i^\pr)|{\cal V}_{_{C}} | e {\overline e} (j^\pr)\rangle\langle
e{\overline e}(j^\pr)|
\;.
\eea
$h$ is the free Hamiltonian given in Eq.~(\ref{eq:hap2}).
We are assuming photons couple perturbatively to the $|e {\overline e} \rangle$ sector,
which must be justified {\em a posteriori}. 
${\cal V}_{_{C}}$
is the Coulomb interaction and will be written explicitly later. Note that 
 the lowest order low-lying spectrum  of the complete $H_{_{R}}$ is
  identical to that of $h+{\cal V}_{_{C}}$
 as long as the limit in Eq.~(\ref{eq:limit}) is taken.

We close this section by writing the standard BSPT 
Raleigh-Schr\"{o}dinger formulae.  For simplicity,
we write the formulae for the non-degenerate case~\cite{baym}:
\bea
\left ( {\cal H}_{o}+{\cal V} \right )
|\Psi_{_{R, N}} ({\cal P}) \rangle &=&
E_{_{ N}} | \Psi_{_{R, N}} ({\cal P}) \rangle~,\\
{\cal H}_{o}  | \psi_{_{ N}} ({\cal P})\rangle &=&
{\cal E}_{_{ N}} | \psi_{_{ N}} ({\cal P}) \rangle~,\\
| \Psi_{_{R, N}} ({\cal P})\rangle&=&
| \psi_{_{ N}} ({\cal P})\rangle + \sum_{_{M \neq N}}
\frac{ | \psi_{_{ M}} ({\cal P})\rangle \frac{\langle
  \psi_{_{M}} ({\cal P})| {\cal V} 
  | \psi_{_{N}} ({\cal P})\rangle}
  {\langle
  \psi_{_{N}} ({\cal P})|  
   \psi_{_{N}} ({\cal P})\rangle}}
   {{\cal E}_{_{N}}-{\cal E}_{_{M}}} +
  { \cal O}({{\cal V}}^{2})~,\\
   E_{_{N}}&=&{\cal E}_{_{N}} +
   \frac{\langle
  \psi_{_{N}} ({\cal P})| {\cal V} 
  | \psi_{_{ N}} ({\cal P})\rangle}
  {\langle
  \psi_{_{N}} ({\cal P})|  
   \psi_{_{N}} ({\cal P})\rangle} +
   \;\;\;\sum_{_{M \neq N}} 
   \frac{\left|\frac{\langle
  \psi_{_{N}} ({\cal P})| {\cal V} 
  | \psi_{_{ M}} ({\cal P})\rangle}
  {\langle
  \psi_{_{N}} ({\cal P})|  
   \psi_{_{N}} ({\cal P})\rangle}\right|^{2}}
   {{\cal E}_{_{N}}-{\cal E}_{_{M}}} +
   {\cal O}({{\cal V}}^{3})~,
\label{eq:hey1}\eea
where ${\cal P}$ is the total three-momentum of the state and ``N" 
labels the total mass of the state.
These general formulae will be used below in \S III to solve
for positronium's spin structure. Note that for the light-front case:
$E_{_{N}}\equiv\frac{{{\cal P}^\perp}^2+M_{_{N}}^2}{{\cal P}^+}$ 
and ${\cal E}_{_{N}}\equiv 
\frac{{{\cal P}^\perp}^2+{\cal M}_{_{N}}^2}{{\cal P}^+}$.
%%%%%%%%%%%%%%%%%%%%%%%%%%%%%%%%%%%%%%%%%%%%%%%%%%%%%%%%%%%%%%%%%%%%%%%%%%%%%
%%%%%%%%%%%%
%%%%%%%%%%%%
%%%%%%%%%%%%                here's chapter 3
%%%%%%%%%%%%
%%%%%%%%%%%%%%%%%%%%%%%%%%%%%%%%%%%%%%%%%%%%%%%%%%%%%%%%%%%%%%%%%%%%%%%%%%%%%%
\vskip.2in
\noindent
\centerline{{\bf III. POSITRONIUM'S SPIN STRUCTURE}}
\vskip.2in
Now we will apply the procedure outlined in \S II to obtain positronium's  
ground state spin splittings   to order $\alpha^4$. We will now give a brief overview of this 
section.
First, we derive $H_{_{R}}$ to second order in $e$.
  This includes a discussion of the effective fermion
self-energy, but the photon self-energy and electromagnetic coupling do not run at this order.   
Then we move on to the diagonalization of $H_{_{R}}$. 
This starts with 
a discussion of our zeroth order Hamiltonian, ${\cal H}_o$, which
will be treated nonperturbatively. This includes a discussion of 
a coordinate
change that takes $(x \in [0,1]) \longrightarrow (\kap_z \in [-\infty,\infty])$,
 which allows easier identification of ${\cal H}_o$. 
 We solve for the spectrum of ${\cal H}_o$ exactly, which among other things,
fixes  the
 $\alpha$-scaling of the momenta in the matrix elements in BSPT.
 Then we move on to 
a derivation of the perturbative effects coming from 
low-energy (energy transfer below $\lam$) photon
emission, absorption and annihilation
 at order $e^2$, which includes a discussion of the full electron and positron self-energies
 and a derivation to order $e^2$ of the complete exchange and annihilation interactions.
 Given this, we determine the 
  range of $\lam$
  that allows the effects of low-energy (energy transfer below $\lam$)
   photon emission and absorption 
 to be  transferred to the effective interactions in the $|e {\overline e}\rangle$
 sector alone, and at the same time,
 does not cut into the nonperturbative features of the
  solutions of ${\cal H}_o$.
 Finally, we  proceed with  BSPT in ${\cal V}$
 noting that all shifts appear in the few-body 
 sector, $|e {\overline e}\rangle$, alone.
%%%%%%%%%%%%%%%%%%%%%%%%%%%%%%%%%%%%%%%%%%%%%%%%%%%%%%%%%%%%%%%%%%%%%%%%%%%%%
%%%%%%%%%%%%
%%%%%%%%%%%%
%%%%%%%%%%%%                here's section 3.1
%%%%%%%%%%%%
%%%%%%%%%%%%%%%%%%%%%%%%%%%%%%%%%%%%%%%%%%%%%%%%%%%%%%%%%%%%%%%%%%%%%%%%%%%%%%
\vskip.2in
\noindent
\centerline{{\bf A. Derivation of {\bm $ H_{_{R}} $ to second order}}}
\vskip.2in

From \S II, the final renormalized Hamiltonian to second order 
is given by:
\bea
\langle i|H_{_{R}}|j\rangle&=& f_{\lam ij} \left\{
h_{ij}+
\left(
v_{can}\right)_{ij}+
\sum_{k} {\left(v_{can}\right)}_{ik} 
  {\left(v_{can}\right)}_{kj} \left(
  \frac{g^{(\lam \Lam)}_{ikj}}{\Delta_{ik}} +
  \frac{g^{(\lam \Lam)}_{jki}}{\Delta_{jk}}
  \right)+
  \del v_{\Lam ij}^{^{(2)}}+{\cal O}(e^3)
\right\}\;.
\label{eq:44}
\eea
$g^{(\lam \Lam)}_{ikj}$ is given in Eq.~(\ref{eq:g}) and $v_{can}$ is given in
Eq.~(\ref{eq:46}).
%%%%%%%%%%%%%%%%%%%%%%%%%%%%%%%%%%%%%%%%%%%%%%%%%%%%%%%%%%%%%%%%%%%%%%%%%%%%%
%%%%%%%%%%%%
%%%%%%%%%%%%
%%%%%%%%%%%%                here's section 3.1.1
%%%%%%%%%%%%
%%%%%%%%%%%%%%%%%%%%%%%%%%%%%%%%%%%%%%%%%%%%%%%%%%%%%%%%%%%%%%%%%%%%%%%%%%%%%%
\vskip.2in
\noindent
\centerline{{\bf 1. Renormalization issues}}
\vskip.2in

The form of $\del v_{_{\Lam}}^{^{(2)}}$
follows from the constraint that $H_{_{R}}$ satisfies 
coupling coherence.
To order $e^2$  the fermion and photon masses run, but the coupling does not. 
First, we  discuss
 the result for the electron
  self-energy coming from the second-order effective interactions in $H_{_{R}}$. We 
 skip the tedious but simple details of the  calculation of the matrix element, 
 but the interested
reader should consult Appendix A, where we have collected our light-front 
conventions for our field expansions and
commutation relations. We use the rules in Ref.~\cite{brent} to calculate our matrix
elements in this work. Specifically,
 we calculate a free matrix element of $H_{_{R}}$ given in Eq.~(\ref{eq:44}) in
 the electron self-energy channel.
   The results are  summarized by the following equation which
  also shows how the
  running electron mass 
  squared, $m_\lam^2$, is defined:\footnote{
 Note that the energy dependence on the 
 electron's relative transverse momentum, $\kap$, does not change.
  }
\bea
\frac{\kappa^2+m_\lam^2}{x}&\equiv&
\frac{\langle e(3) {\overline e}(4) |
H_{_{R}} | e(1) {\overline e}(2)\rangle|_{_{self \;energy}}\; {\cal P}^+}
{\langle e(3) {\overline e}(4) |
 e(1) {\overline e}(2)\rangle} -{{\cal P}^\perp}^2
 -\frac{\kappa^2+m^2}{1-x}\nonumber\\
 &=&
\frac {\kappa^2+m^2}{x}+\frac{-(\del\Sigma_\Lam^{^{(2)}}
-\del\Sigma_\lam^{^{(2)}})+\del v_{\Lam}^{^{(2)}}}{x}+{\cal O}(e^4)
\label{eq:72}\;,
\eea
 where 
 \bea
 \del\Sigma_{\lam^\pr}^{^{(2)}}&=& \frac{\alpha}{2 \pi} 
 \left(
 -\frac{3 {\lam^\pr}^2 x+m^2}{2}
 +\frac{m^2 \left(\frac{m^2}{2}+{\lam^\pr}^2 x
 \right)}{m^2+{\lam^\pr}^2 x}
 -3 m^2 \log \left(\frac{m^2+{\lam^\pr}^2 x}{m^2}
 \right)\right) \nonumber\\
&&~~~~+\frac{\alpha}{2 \pi} 
 \left(-2 {\lam^\pr}^2 x\log \left(\frac{m^2+{\lam^\pr}^2 x}{{\lam^\pr}^2 x}
 \right)
 +2 {\lam^\pr}^2 x \log\left(\frac{ x}{\eps}
 \right)
 \right)
 \;.
 \label{eq:fermionmass}
\eea
In the respective $\del\Sigma_{\lam^\pr}^{^{(2)}}$ terms of Eq.~(\ref{eq:72}),
 ${\lam^\pr}=\Lam$ and $\lam$.
In these formulae,  
\beaa
&& p_{_{electron}} = p_{_{1}}=
 (x {\cal P}^+, \kappa +~x~{\cal P}^\perp)~~{\rm and}~~
{\cal P}_{_{positronium}}={\cal P}=({\cal P}^+,{\cal P}^\perp)~,
\eeaa
   and $|e(1)\rangle$ 
(~or $|{\overline e}(1)\rangle$~) is a 
state of the free Hamiltonian, $h$, with
spin and momenta coordinates labeled by ``1."
Note that we are forced
 to introduce an infrared regulator, $\eps$. This is introduced by requiring all
longitudinal momenta (real, virtual or instantaneous)  to satisfy:
\bea
\frac{|p^+|}{{\cal P}^+} &\geq& \eps=0_{+}\;,
\label{eq:ir}
\eea
where ${\cal P}^+$ is the total longitudinal momentum of the physical state.
The absolute value sign is required for  instantaneous lines. 
 Physical results can not depend on this infrared regulator, $\eps$, and in this QED 
calculation we show that treating the photon perturbatively leads to an exact cancellation of
this infrared divergence in the above running electron mass squared, $m_\lam^2$.

We constrain the electron mass to run coherently with the cutoff which from 
Eq.~(\ref{eq:cc}) and Eq.~(\ref{eq:72}) amounts to the requirement:
\bea
m_\lam^2&=&\left[m^2-(\del\Sigma_\Lam^{^{(2)}}
-\del\Sigma_\lam^{^{(2)}})+\del v_{\Lam}^{^{(2)}} +{\cal O}\left(e^4\right)\right]
=\left[m^2+\del v_{\Lam}^{^{(2)}}+{\cal O}\left(e^4\right)\right]_{\Lam \rightarrow \lam}
\;.\label{eq:1945}
\eea
This fixes the mass counterterm:\footnote{Actually, any finite ${\cal O}(e^2)$ {\em scale independent}
term could be added to the counterterm, and Eq.~(\ref{eq:1945}) would still be satisfied. However,
 higher order renormalization reveals that this term would not be coherent, and so is excluded.}
\bea
\del v_{\Lam}^{^{(2)}}&=&\del\Sigma_{\Lam}^{^{(2)}}+{\cal O}(e^4)\;,
\label{eq:1000}
\eea
and to second order the fermion mass renormalization is complete.
 
For arbitrary $\lam$, 
the photon mass also runs at order $e^2$. The discussion follows that of the electron mass except 
for the fact that the running photon mass is infrared finite. 
For $\lam^2 < (2 m)^2$, the photon mass does not run because pair production is no longer possible.
Thus, for $\lam^2 \ll m^2$ the photon mass is zero to all orders in perturbation theory.
There are additional difficulties with marginal operators that are encountered
at ${\cal O}(e^3)$, but this is beyond the focus of this paper.
%%%%%%%%%%%%%%%%%%%%%%%%%%%%%%%%%%%%%%%%%%%%%%%%%%%%%%%%%%%%%%%%%%%%%%%%%%%%%
%%%%%%%%%%%%
%%%%%%%%%%%%
%%%%%%%%%%%%                here's section 3.1.2
%%%%%%%%%%%%
%%%%%%%%%%%%%%%%%%%%%%%%%%%%%%%%%%%%%%%%%%%%%%%%%%%%%%%%%%%%%%%%%%%%%%%%%%%%%%
\vskip.2in
\noindent
\centerline{{\bf 2. {\bm $ H_{_{R}} $} to order {\bm $e^2$}: exchange and annihilation channels}}
\vskip.2in

To complete the derivation of $H_{_{R}}$ to second order we need to write the
coherent interactions for the exchange and annihilation channels in the $|e {\overline e} \rangle$
sector.
At second order, these come from
 tree level diagrams, with no divergences or running couplings, thus the results
follow from:
\bea
\del v_{\Lam}^{^{(2)}}&\equiv& -\int_\Lam^\infty [ v_{\lam^\pr}^{^{(1)}},T_{\lam^\pr}^{^{(1)}}]
d \lam^\pr
\;.
\label{eq:iiii}
\eea
To show that 
$\del v_{\Lam}^{^{(2)}}$ produces a coherent interaction recall Eq.~(\ref{eq:23}). We have:
\bea
{\overline v}_\lam &=& v_{can}-\int_\Lam^\infty [ v_{\lam^\pr}^{^{(1)}},T_{\lam^\pr}^{^{(1)}}]
d \lam^\pr -\int_\lam^\Lam [ v_{\lam^\pr}^{^{(1)}},T_{\lam^\pr}^{^{(1)}}]
d \lam^\pr\nonumber +{\cal O}(e^3)\\
&=&v_{can}-\int_\lam^\infty [ v_{\lam^\pr}^{^{(1)}},T_{\lam^\pr}^{^{(1)}}]
d \lam^\pr+{\cal O}(e^3)\\
&=&\left.\left[v_{can}-\int_\Lam^\infty [ v_{\lam^\pr}^{^{(1)}},T_{\lam^\pr}^{^{(1)}}]
d \lam^\pr+{\cal O}(e^3)
\right]\right|_{\Lam \rightarrow \lam}
\;,
\eea
which satisfies the coupling coherence constraint, Eq.~(\ref{eq:cc}). 
At second order this seems trivial, but at higher orders the constraint that only
$e$ and $m$ run independently with the cutoff places severe constraints on the Hamiltonian.

Given this second order interaction, the free matrix elements of $H_{_{R}}$, shown  
 in Eq.~(\ref{eq:44}), in the exchange and annihilation channels 
are:

\noindent
\underline{Exchange Channel}
 \bea
 V_{_{\lam,exchange}}&\equiv&\frac{\langle e (3) {\overline e} (4)|
 H_{_{R}}|e(1){\overline e}(2)\rangle|_{_{exchange}}}
 {16\pi^3\del^3(p_1+p_2-p_3-p_4) \sqrt{x x^\prime (1-x)(1-x^\prime)}}\nonumber\\
 &\equiv&
 V_1+V_2
 +{\cal O}(e^4)\;,
 \label{eq:saeko3}
 \eea 
 where
 \bea
V_1&=&- \;e^2 \;N_1 \;\;\theta\left(
 \lam^2-\left|{\cal M}_o^2-
 {{\cal M}_o^\pr}^2\right|\right)\nonumber\\
&&~~~~\times \left(
  \frac{\theta\left(
 \left|\Delta_1\right|-\left|\Delta_2\right|\right)
 \theta\left(
 \left|\Delta_1\right|-\lam^2\right)}{DEN_1}
 +
  \frac{\theta\left(
 \left|\Delta_2\right|-\left|\Delta_1\right|\right)
 \theta\left(
 \left|\Delta_2\right|-\lam^2\right)}{DEN_2}
  \right)\;,\\
V_2&=&- \;e^2\;\theta\left(
 \lam^2-\left|{\cal M}_o^2-
 {{\cal M}_o^\pr}^2\right|\right)
 \left(
 \frac{4 }{(x-x^\prime)^2}\del_{s_1s_3}\del_{s_2s_4}
 \right)
\;.
\label{eq:76}
\eea
The variables  are defined as follows (see Figure~1 also):
\beaa
\bullet\;\; p_1&=&\left(x {\cal P}^+ , \kappa+x {\cal P}^\perp\right)\;,
\;p_2=\left((1-x) {\cal P}^+ , -\kappa+(1-x) {\cal P}^\perp\right)\;\\
\bullet\;\; p_3&=&\left(x^\prime 
{\cal P}^+ , \kappa^\prime+x^\prime {\cal P}^\perp\right)\;,
\;p_4=\left((1-x^\prime) {\cal P}^+ , -\kappa^\prime+(1-x^\prime)
 {\cal P}^\perp\right)\nonumber\;
\\
\bullet\;\; N_1&=&\del_{s_1 s_3}\del_{s_2 s_4} T_1^\perp \cdot T_2^\perp-
2 m^2 \del_{s_2 {\overline s_4}}\del_{s_2 {\overline s_1}}
\del_{s_3 {\overline s_1}}\frac{(x-x^\prime)^2}{x 
x^\prime (1-x) (1-x^\prime)}\\
&+&i m \sqrt{2} (x^\prime-x)\left(
\frac{s_1}{x x^\prime}\del_{{\overline s_1} s_3}\del_{s_2 s_4} 
\eps_{s_1}^\perp \cdot T_1^\perp+
\frac{s_2}{(1-x) (1-x^\prime)}\del_{{\overline s_4} s_2}\del_{s_1 s_3} 
\eps_{s_2}^\perp \cdot T_2^\perp
\right)\;\\
\bullet\;\;T_1^i&=&-\frac{2 (\kappa^i-{\kappa^\prime}^i)}{x-x^\prime}
-\frac{{\kappa^\prime}^i({\overline s_2})}{1-x^\prime}
-\frac{\kappa^i(s_2)}{1-x}\;,\;
T_2^i=\frac{2 (\kappa^i-{\kappa^\prime}^i)}{x-x^\prime}
-\frac{{\kappa^\prime}^i({\overline s_1})}{x^\prime}
-\frac{\kappa^i(s_1)}{x}\\
\bullet\;\;\kappa^i(s)&=&\kappa^i+i \;s \;\eps_{ij}\; \kappa^j
\;\;(s\;=\;\pm\;1\;{\rm and}\;{\overline s}\equiv -s)\;\;;
\eps_{12}=-\eps_{21}=1\;,\;\eps_{11}=\eps_{22}=0\\
\bullet\;\;\Delta_1&=&\frac{DEN_1}{x^\pr-x}\;,\;\Delta_2=\frac{DEN_2}{x^\pr-x}\\
\bullet\;\;DEN_1&=&\frac{(\kap x^\prime-\kap^\prime x)^2}{x x^\prime}+
\frac{(m x-m x^\pr)^2}{x x^\pr}\;,\;
DEN_2=DEN_1|_{x \rightarrow 1-x\;,\;x^\pr \rightarrow 1-x^\pr}\\
\bullet\;\;{\cal M}_o^2&=&\frac{{\kap}^2+m^2}{x (1-x)}\;,\;
{{\cal M}_o^\pr}^2=\frac{{\kap^\pr}^2+m^2}{x^\pr (1-x^\pr)}\;.
\eeaa

\noindent
\underline{Annihilation Channel}
\bea
 V_{_{\lam,annihil}}&\equiv&\frac{\langle e (3) {\overline e} (4)|
 H_{_{R}}|e(1){\overline e}(2)\rangle|_{_{annihilation}}}
 {16\pi^3\del^3(p_1+p_2-p_3-p_4) \sqrt{x x^\prime (1-x)(1-x^\prime)}}\nonumber\\
 &\equiv&
 V_3+
 V_4
 +{\cal O}(e^4)\;,
 \label{eq:saeko4}
 \eea 
 where
 \bea
&&V_3= \;e^2 \;N_2 \;\;\theta\left(
 \lam^2-\left|{\cal M}_o^2
-{{\cal M}_o^\pr}^2\right|\right)
 \left(
  \frac{\theta\left({\cal M}_o^2
-{{\cal M}_o^\pr}^2
 \right)
 \theta\left(
 {\cal M}_o^2-\lam^2\right)}{{\cal M}_o^2}\right.\nonumber\\
 &&~~~~~~~~~~~~~~~~~~~~~~~~~~~~~~~~~~~~~~~~~~~~~~~+\left.
  \frac{\theta\left({{\cal M}_o^\pr}^2-{\cal M}_o^2
 \right)
 \theta\left(
 {{\cal M}_o^\pr}^2-\lam^2\right)}{{{\cal M}_o^\pr}^2}
  \right)~,\\
&&V_4=\;4 e^2 \theta\left(
 \lam^2-\left|{\cal M}_o^2
-{{\cal M}_o^\pr}^2\right|\right)
\del_{s_1 {\overline s_2}} \del_{s_3 {\overline s_4}}\;,\label{eq:747}
\eea
and
\beaa
\;\; N_2&=&\del_{s_1 {\overline s_2}}\del_{s_4 {\overline s_3}} T_3^\perp 
\cdot T_4^\perp+
 \del_{s_1  s_2}\del_{s_1  s_4}
\del_{s_3  s_4}\frac{2 m^2}{x 
x^\prime (1-x) (1-x^\prime)}\\
&&~~~~+ i m \sqrt{2} \left(
\frac{s_1}{x (1-x)} \del_{s_1 s_2}\del_{ s_4 {\overline s_3}} 
\eps_{s_1}^\perp \cdot T_3^\perp-
\frac{s_4}{x^\prime (1-x^\prime)}\del_{ s_1 {\overline s_2}}\del_{ s_3 s_4} 
{\eps_{s_4}^\perp}^\ast \cdot T_4^\perp
\right)\;,\\
\;\;T_3^i&=&
\frac{{\kappa^\prime}^i( s_3)}{1-x^\prime}
-\frac{{\kappa^\pr}^i({\overline s_3})}{x^\pr}\;,\;
T_4^i=
\frac{{\kappa}^i({\overline s_1})}{1-x}-\frac{{\kappa}^i( s_1)}{x}
\;.\eeaa
$V_2$ and $V_4$ are  canonical instantaneous exchange and annihilation
interactions, respectively, with widths restricted by the
regulating function, $f_\lam$. 
$V_1$ and $V_3$ are effective  interactions that arise because 
 photon emission and annihilation have vertices with widths restricted 
by the
regulating function, $f_\lam$. 
%%%%%%%%%%%%%%%%%%%%%%%%%%%%%%%%%%%%%%%%%%%%%%%%%%%%%%%%%%%%%%%%%%%%%%%%%%%%%
%%%%%%%%%%%%
%%%%%%%%%%%%
%%%%%%%%%%%%                 section 3.2
%%%%%%%%%%%%                        
%%%%%%%%%%%%
%%%%%%%%%%%%%%%%%%%%%%%%%%%%%%%%%%%%%%%%%%%%%%%%%%%%%%%%%%%%%%%%%%%%%%%%%%%%%%
\vskip.2in
\centerline{{\bf B. Diagonalization of {\bm $H_{_{R}}$}}}
\vskip.2in

First we discuss the lowest order spectrum of $H_{_{R}}$, after which we 
 discuss BSPT, renormalization and a limiting procedure which allows
 the effects of low-energy (energy transfer below $\lam$)
 emission to be transferred to the $|e {\overline e}\rangle$
 sector alone.
%%%%%%%%%%%%%%%%%%%%%%%%%%%%%%%%%%%%%%%%%%%%%%%%%%%%%%%%%%%%%%%%%%%%%%%%%%%%%
%%%%%%%%%%%%
%%%%%%%%%%%%
%%%%%%%%%%%%                here's section 3.2.1
%%%%%%%%%%%%
%%%%%%%%%%%%%%%%%%%%%%%%%%%%%%%%%%%%%%%%%%%%%%%%%%%%%%%%%%%%%%%%%%%%%%%%%%%%%%
\vskip.2in
\centerline{{\bf 1. {\bm ${\cal H}_o$}, a coordinate change and its exact spectrum}}
\vskip.2in

 ${\cal H}_o$ in the $|e {\overline e}\rangle$
 sector is
\bea
{\cal H}_o&=&h+{\cal V}_{_{C}}\;,
\eea
where $h$ is the free Hamiltonian given in Eq.~(\ref{eq:hap2}),
and ${\cal V}_{_{C}}$ 
is given by (using the same variables defined below Eq.~(\ref{eq:76});
note,
$\kap_z$ is defined below by Eq.~(\ref{eq:xonehalf}))
\bea
V_{_{C}}&\equiv&\frac{\langle e (3) {\overline e} (4)|
 {\cal V}_{_{C}}|e(1){\overline e}(2)\rangle}
 {16\pi^3\del^3(p_1+p_2-p_3-p_4) \sqrt{x x^\prime (1-x)(1-x^\prime)}}\equiv
 -\frac{16 m^2 e^2 \del_{s_1 s_3} \del_{s_2 s_4}}{(\kap-\kap^\pr)^2+(\kap_z-\kap_z^\pr)^2}
\;.\label{eq:ruo}
\eea
In all other sectors we choose ${\cal H}_o = h$. ${\cal H}_o$ 
 in the $|e{\overline e}\rangle$ sector was motivated from the form of our second order renormalized
  Hamiltonian,
 $H_{_{R}}$, and arises from a nonrelativistic limit of the
  instantaneous photon exchange interaction
 combined with the two time orderings of the dynamical photon exchange interaction.
 We choose it to simplify positronium bound-state calculations. Other choices are possible, and must
 be used to study problems such as photon emission.
 Later, in BSPT this choice is shown to produce the leading order contribution to 
 positronium's mass
 as long as the limit in Eq~(\ref{eq:limit}) is taken.
 
 The coordinates
 $\kap_z$ and $\kap_z^\pr$ in Eq.~(\ref{eq:ruo}) 
 follow from a standard coordinate transformation that
  takes
 the range of longitudinal momentum fraction, $x \in [0,1]$ to 
 $\kap_z \in [-\infty, \infty]$.
 This coordinate change is:
 \bea
 x&\equiv&\frac{1}{2}+\frac{\kap_z}{2 \sqrt{\kap^2+\kap_z^2+m^2}}\;.
 \label{eq:xonehalf}\eea
 We introduce  
 a new three-vector defined as:
  \bea
 {\bf p}&\equiv& (\kap,\kap_z)
 \;.
 \label{eq:nicerelation}\eea
 Note that
 \bea
 {\cal M}_o^2&\equiv&\frac{\kap^2+m^2}{x (1-x)}=4(m^2+{\bf p}^2)
 \eea
 is invariant with respect to rotations in the space of vectors ${\bf p}$.
The nonrelativistic assumption of Eqs.~(\ref{eq:nr1}) and (\ref{eq:nr2}) in 
terms of this  three-vector
becomes:
\bea
\frac{|{\bf p}|}{m} &=& {\cal O}(\alpha)\;.\label{eq:nr3}
\eea

Note
 the  simple forms that our  ``exchange channel denominators" take in the
 nonrelativistic limit:
 \bea
 DEN_1&=& ({\bf p}-{\bf p}^\pr)^2
 -\frac{(\kap_z-\kap_z^\pr) ({\bf p}^2-{{\bf p}^\pr}^2)}{m}
 +{\cal O}\left[\left(\frac{{\bf p}}{m}\right)^5 m^2\right]\label{eq:ohno}~,\\
 DEN_2&=& ({\bf p}-{\bf p}^\pr)^2
 +\frac{(\kap_z-\kap_z^\pr) ({\bf p}^2-{{\bf p}^\pr}^2)}{m}
 +{\cal O}\left[\left(\frac{{\bf p}}{m}\right)^5 m^2\right]~.\label{eq:ohno2}
 \eea
 Also note the form that the longitudinal momentum fraction 
 transferred between the electron and positron
 takes:
 \bea
 x-x^\pr &=& \frac{\kap_z-\kap_z^\pr}{2 m}+\frac{({{\bf p}^\pr}^2 \kap_z^\pr-{\bf p}^2 \kap_z)}{4 m^3}
 +{\cal O}\left[\left(\frac{{\bf p}}{m}\right)^5\right]
 \;.
 \label{eq:96}\eea
 These formulae are used throughout this paper.
  
  Now we describe the leading 
  Schr{\"o}dinger equation. We seek  solutions
  of the following eigenvalue equation:
\bea
{\cal H}_o |\psi_N({\cal P})\rangle&=&{\cal E}_N |\psi_N({\cal P})\rangle\;,
\label{eq:3333}
\eea
where ${\cal E}_N\equiv \frac{{{\cal P}^\perp}^2+{\cal M}_N^2}{{\cal P}^+}$.
 {\em ${\cal H}_o$  is diagonal with respect to the different particle sectors, thus
we can solve Eq.~(\ref{eq:3333}) sector by sector.}
In all sectors other than  $|e {\overline e}\rangle$,
${\cal H}_o=h$, and the solution is trivial.
For the $|e {\overline e}\rangle$ sector,
a general $|\psi_N({\cal P})\rangle$ is:
 \bea
 |\psi_N({\cal P})\rangle&=&\sum_{s_1 s_2} \int_{p_1 p_2} \sqrt{p_1^+ p_2^+}
 16 \pi^3 \del^3({\cal P}-p_1-p_2) \;{\tilde{\phi}}_N(x \kap s_1 s_2)\;
 b_{s_1}^\dagger(p_1)\; d_{s_2}^\dagger(p_2)\; |0\rangle\;,
 \eea
 with norm:
 \beaa
 &&\langle \psi_{N} ({\cal P})|\psi_{N^\pr} ({\cal P}^\pr)\rangle \equiv
 \del_{N N^\pr}16\pi^3 {\cal P}^+
 \del^3\left({\cal P}-{\cal P}^\pr\right)\\
 &&\Longrightarrow\;
 \sum_{s_1 s_2}\frac{\int d^2 \kap \int_0^1 dx}{16\pi^3}
 {{\tilde{\phi}}_N}^\ast(x \kap s_1 s_2) {\tilde{\phi}}_{N^\pr}(x \kap s_1 s_2)\;=\;\del_{N N^\pr}\;.
 \eeaa
 The tilde on ${\tilde{\phi}}_{N}$ will be notationally convenient below. 
 In the $|e {\overline e}\rangle$ sector, Eq.~(\ref{eq:3333}) becomes:
 \bea
 \left({\cal M}_{_{N}}^2-\frac{{\kap^\pr}^2+m^2}{x^\pr (1-x^\pr)}
 \right) {\tilde{\phi}}_N(x^\pr \kap^\pr s_3 s_4)&=&
 \sum_{s_1 s_2}\frac{\int d^2 \kap \int_0^1 dx}{16\pi^3}
 ~V_{_{C}}~ {\tilde{\phi}}_N(x \kap s_1 s_2)
 \;.\label{eq:sc}
 \eea
 After the above coordinate change, this becomes:
 \bea
 \left({\cal M}_{_{N}}^2-4 (m^2+{{\bf p}^\pr}^2)
 \right) \phi_N({\bf p}^\pr  s_3 s_4)&=&
 \sum_{s_1 s_2}\int\frac{ d^3 p \sqrt{J(p)J(p^\prime)}}{16\pi^3}
 ~V_{_{C}}~
  \phi_N({\bf p} s_1 s_2)
 \;,\label{eq:full}
 \eea
 where the tilde on the wavefunction has been removed by redefining the norm
 in a convenient fashion:
 \bea
\del_{N N^\pr}&=&\sum_{s_1 s_2}\frac{\int d^2 \kap \int_0^1 dx}{16\pi^3}
 {\tp_N}^\ast(x \kap s_1 s_2) \tp_{N^\pr}(x \kap s_1 s_2) =
 \sum_{s_1 s_2} \int d^3 p \frac{J(p)}{16\pi^3}
 {\tp_N}^\ast({\bf p} s_1 s_2) \tp_{N^\pr}({\bf p} s_1 s_2)
 \nonumber\\
 &\equiv&\sum_{s_1 s_2} \int d^3 p \;\phi_N^{ \ast}({\bf p} s_1 s_2) \phi_{N^\pr}({\bf p} s_1 s_2)
 \;,\label{eq:norm}\eea
 and the Jacobian of the transformation is given by:
 \bea
 J(p)&\equiv&\frac{dx}{d\kap_z}=\frac{\kap^2+m^2}{2({\bf p}^2+m^2)^{\frac{3}{2}}}
 \;.\eea
 Note that the Jacobian factor in Eq.~(\ref{eq:full}) satisfies:
 \bea
 \sqrt{J(p)J(p^\prime)}&=& \frac{1}{2 m}\left(1-
\frac{{\bf p}^2+2\kap_z^2+{{\bf p}^\prime}^2+2{\kap_z^\pr}^2}{4 m^2}+
{\cal O}\left(\frac{{\bf p}^4}{m^4},\frac{{{\bf p}^\pr}^4}{m^4},\cdots\right)\right)
 \;.\label{eq:jac}
 \eea
 
 Before defining ${\cal H}_o$ in the $|e {\overline e}\rangle$ sector 
 we mention a subtle but important point in 
 the definition of ${\cal H}_o$.  {\em
 ${\cal H}_o$ in the $|e {\overline e}\rangle$ sector 
 will not be defined by Eq.~(\ref{eq:full})}. Rather, it will be defined by
 taking the leading order
  nonrelativistic expansion of the Jacobian factor in Eq.~(\ref{eq:full}). This gives
   \bea
 \left({\cal M}_{_{N}}^2-4 (m^2+{{\bf p}^\pr}^2)
 \right) \phi_N({\bf p}^\pr  s_3 s_4)&=&
 \sum_{s_1 s_2}\int\frac{ d^3 p \left(\frac{1}{2 m}\right)}{16\pi^3}
 ~V_{_{C}}~
  \phi_N({\bf p} s_1 s_2)
 \;,
 \label{eq:cccoul}
 \eea
 where $V_{_{C}}$ is defined in Eq.~(\ref{eq:ruo}). This ${\cal H}_o$ will be diagonalized
 exactly, and the subsequent BSPT will be set up as an expansion in
 ${\cal V}\equiv H_{_{R}}-{\cal H}_o$. First, we discuss the exact diagonalization of
 ${\cal H}_o$.
 
 Putting  the expression for $V_{_{C}}$ into Eq.~(\ref{eq:cccoul}) results in the following
 equation:
 \bea
 \left(-{\cal B}_{_{N}}+\frac{{{\bf p}^\pr}^2}{m}
 \right) \phi_N({\bf p}^\pr  s_3 s_4)&=&
 \frac{\alpha}{2 \pi^2}\int\frac{ d^3 p }{({\bf p}-{\bf p}^\pr)^2}
  \phi_N({\bf p} s_3 s_4)
 \;.\label{eq:C}
\eea
 This is recognized as the familiar nonrelativistic Schr{\"o}dinger equation for positronium.
 Note that we have defined a leading order binding energy, ${\cal B}_N$, as:
 \bea
 {\cal M}_N^2 &\equiv& 
 4 m^2 + 4 m {\cal B}_{_{N}}
 \;.
 \eea
 Note the difference in the definition of this leading order binding energy and the full binding 
 energy as given by $M_{_{N}}^2\equiv(2 m+B_{_{N}})^2$ (see Appendix D for further discussion
 of this difference).
 
 To proceed with the solution of Eq.~(\ref{eq:C}) note that there is no spin dependence in
 the operator so the spin part just factors out:
  \bea
\phi_{\mu, s_e,s_{_{{\overline e}}}}({\bf p}^\pr s_3s_4)&\equiv&\phi_\mu({\bf p}^\pr) 
\del_{s_e s_3} \del_{s_{_{{\overline e}}} s_4}
\;.
\eea
 We rewrote
$N$ as $(\mu, s_e,s_{_{{\overline e}}})$, where $(s_e,s_{_{{\overline e}}})$  label the
spin quantum numbers and $\mu$ labels all other quantum numbers, which are discrete for the bound states
and continuous for the scattering states.

The solutions to Eq.~(\ref{eq:C}) are well known.
 For ${\cal B}_{_{N}} < 0$, following Fock~\cite{fock1935}, 
 we change coordinates according to
 \bea
 m {\cal B}_{_{N}}&\equiv&-e_n^2~,\\
u&\equiv&(u_0,{\bf u}) ~,\\
 u_0&\equiv&\cos(\omega)\equiv\frac{e_n^2-{\bf p}^2}{e_n^2+{\bf p}^2}~,\\
 {\bf u}&\equiv&\frac{{\bf p}}{p}\sin(\omega) \equiv \sin(\omega)\left(
\sin(\theta) \cos(\phi) ,sin(\theta) \sin(\phi) , \cos(\theta)\right)\nonumber\\
&\equiv& \frac{2 e_n {\bf p}}{e_n^2+{\bf p}^2}
 \;.\eea
 Useful relations implied by this coordinate change are in Appendix~B.
 Note that in our  notation we anticipate that $\mu$ will be given 
 by $(n, l, m_{_{l}})$, the usual principal and angular momentum
  quantum numbers, and that the leading order binding
 will  depend only on the principal quantum number, $n$.
Given this, 
 Eq.~(\ref{eq:C}) becomes
\bea
\psi_\mu(\Omega^\pr)&\equiv&\frac{\alpha}{2 \pi^2}\frac{m}{2 e_n} 
\int  \frac{d \Omega}{|u-u^\pr|^2} \psi_\mu(\Omega)
\;,\eea
where
\bea
\psi_\mu(\Omega)&\equiv&\frac{(e_n^2+{\bf p}^2)^2}{4 (e_n)^{\frac{5}{2}}} \phi_\mu({\bf p})\;.
\label{eq:110}
\eea
Using Eq.~(\ref{eq:1900}) of
 Appendix B, this is seen to have the following solution:
\bea
&&\psi_\mu(\Omega)=Y_\mu(\Omega)~~{\rm with}~~ 
\frac{\alpha}{2 \pi^2}\frac{m}{2 e_n} \frac{2 \pi^2}{n} = 1
\;,\eea
where
$Y_\mu(\Omega)$ is a hyperspherical harmonic.
Thus,
\bea
e_n=\frac{m \alpha}{2 n}~~{\rm and}~~
 {\cal B}_{_{N}}=-\frac{m \alpha^2}{4 n^2}
\;.
\label{eq:112}
\eea
  This 
is the standard nonrelativistic solution for the bound states of positronium
to order $\alpha^2$.
This completes the solution
of ${{\cal H}}_o$ for the bound states. 
The scattering $|e{\overline e}\rangle$
  states are also needed in our BSPT calculation.
We  use propagator techniques to include these scattering
states where required (see Appendix~C).

%%%%%%%%%%%%%%%%%%%%%%%%%%%%%%%%%%%%%%%%%%%%%%%%%%%%%%%%%%%%%%%%%%
%%%%%%%%%%%%
%%%%%%%%%%%%
%%%%%%%%%%%%                here's section 3.2.2
%%%%%%%%%%%%
%%%%%%%%%%%%%%%%%%%%%%%%%%%%%%%%%%%%%%%%%%%%%%%%%%%%%%%%%%%%%%%%%%%%%%%%%%%%%%
\vskip.2in
\noindent
\centerline{{\bf 2. BSPT, renormalization and a limit}}
\vskip.2in

Here we use the BSPT formulae (appropriately generalized to the degenerate case) of \S II.B 
to analyze positronium's spin structure.
The potential to be used in BSPT is:
\bea
{\cal V}&=&H_{_{R}}-{\cal H}_o\;,
\eea
where the eigenvalue equation for ${\cal H}_o$  is given by Eq.~(\ref{eq:cccoul}),
and $H_{_{R}}$ to second order is given in \S III.A. We will be perturbing about 
the nonperturbative eigenstates
 of ${\cal H}_o$.  

First, we discuss electron mass renormalization. In second order BSPT there is an
electron mass shift coming from the $f_\lam v_{_{can}}$ part of $H_{_{R}}$, with
$v_{_{can}}$ given by $\int d^2 x^\perp d x^- {\cal H}_{e e \gamma}$ (see Eq.~(\ref{eq:48})).
This is photon emission and absorption restricted by the regulating function, $f_\lam$.
The calculation is similar to that of \S III.A.1. Assuming
$\langle {\cal M}_{_{N}}^2-{\cal M}_o^2\rangle = {\cal O}(e^2)$, this
electron mass-squared shift is
\bea
\del m^2&=&-\del\Sigma_\lam^{^{(2)}}+{\cal O}(e^4)\;.
\label{eq:35}
\eea
$\del\Sigma_\lam^{^{(2)}}$ is the same function that was defined in Eq.~(\ref{eq:fermionmass}).
Using this result (Eq.~(\ref{eq:35})) 
one obtains: $\langle {\cal M}_{_{N}}^2-{\cal M}_o^2\rangle
=\langle 4 m^2+ 4 m {\cal B}_{_{N}}-4 (m^2+{\bf p}^2)\rangle={\cal O}(e^4)$,
and our initial assumption is satisfied. When
this is combined with the only other second order electron 
mass shift, $m_\lam^2$, of Eq.~(\ref{eq:72}) 
we have, for the full electron mass-squared, to 
second order,
\bea
m_e^2&=&m_\lam^2+\del m^2\nonumber\\
&=& \left[m^2-\left(\del\Sigma_{_{\Lam}}^{^{(2)}}- 
\del\Sigma_{_{\lam}}^{^{(2)}}\right)+\del v_{_{\Lam}}^{^{(2)}}\right]+\left[-
\del\Sigma_{_{\lam}}^{^{(2)}}\right]+{\cal O}(e^4)\nonumber\\
&=&m^2+{\cal O}(e^4)
\;.
\eea
In this last step we recalled the result from Eq.~(\ref{eq:1000}).
We see that to second order, the full electron mass is given by the electron
mass in the free Hamiltonian, $h$. 
Also, as promised below Eq.~(\ref{eq:ir}), we see that treating photons perturbatively
has led to an exact cancellation of the infrared divergence in the
running mass, $m_\lam^2$:  the full electron mass,
$m_e^2$, to second order is infrared finite.

Now we move on to the discussion of BSPT.
The only  channels to order $e^2$ are exchange and annihilation. Parts of these effective
interactions are given in \S III.A.2 . We also need to include the perturbative
mixing of the $|e {\overline e} \gamma \rangle$ 
and $|\gamma\rangle$ sectors with the 
$|e {\overline e}  \rangle$ sector arising from $f_\lam v_{_{can}}$, with 
$v_{_{can}}=\int d^2 x^\perp d x^- {\cal H}_{e e \gamma}$. In second
order BSPT this gives rise to the following effective interactions that must be added to 
$V_{_{\lam,exchange}}$ and $V_{_{\lam,annihil}}$ 
of Eqs.~(\ref{eq:saeko3}) and (\ref{eq:saeko4})
respectively.

\noindent
\underline{Exchange Channel}
\bea
V_5&=& \frac{- e^2 N_1
\theta\left(\lam^2-\left|\Delta_1\right|\right)
\theta\left(\lam^2-\left|\Delta_2\right|\right)}{DEN_3}
\;,
\eea
with 
\beaa
DEN_3 &=& (\kap -\kap^\pr)^2+\frac{1}{2} (x-x^\pr) A +|x-x^\pr|
\left( \frac{1}{2} \left( {\cal M}_o^2+{{\cal M}_o^\pr}^2 \right) -{\cal M}_{_{N}}^2\right)
\;,\eeaa
and
\beaa
A&=& \frac{\kap^2+m^2}{1-x}-\frac{{\kap^\pr}^2+m^2}{1-x^\pr}+\frac{{\kap^\pr}^2+m^2}{x^\pr}
-\frac{\kap^2+m^2}{x}
\;.
\eeaa

\noindent
\underline{Annihilation Channel}
\bea
V_6&=&\;e^2 \;N_2 \;\;\frac{\theta\left(\lam^2-{\cal M}_o^2\right)
\theta\left(\lam^2-{{\cal M}_o^\pr}^2\right)}{{\cal M}_{_{N}}^2}
\;.
\eea
Note that in a nonrelativistic expansion (after the coordinate change of Eq.~(\ref{eq:xonehalf})),
the above ``exchange channel denominator" becomes
\bea
DEN_3&=&({\bf p}-{\bf p}^\pr)^2+|x-x^\pr|
\left( \frac{1}{2} \left( {\cal M}_o^2+{{\cal M}_o^\pr}^2 \right) -{\cal M}_{_{N}}^2\right)+
 {\cal O}\left[\left(\frac{{\bf p}}{m}\right)^6 m^2\right]
\;.
\eea

The full exchange and annihilation channel interactions to order $e^2$ are
\bea
V_{_{exchange}}&\equiv&V_{_{\lam,exchange}}+V_5\label{eq:saeko1}\;,\\
V_{_{annihil}}&\equiv&V_{_{\lam,annihil}}+V_6\label{eq:saeko2}
\;,
\eea
where
 Eqs.~(\ref{eq:saeko3}) and (\ref{eq:saeko4}) give $V_{_{\lam,exchange}}$ and 
$V_{_{\lam,annihil}}$ respectively.

One way to summarize the results, recalling the form of Eq.~(\ref{eq:full})
and the norm in Eq.~(\ref{eq:norm}), is to state:
the full order $e^2$ effective interactions give rise to the following first order BSPT shift of the 
 bound-state mass-squared spectrum of ${\cal H}_o$:
\bea
\del^{^{(1)}}M^2(s_3,s_4;s_1,s_2)
&\equiv&\langle \phi_{n,l,m_l,s_3,s_4}|V|\phi_{n,l,m_l,s_1,s_2}\rangle\nonumber\\
&=& \int d^3 p~ d^3 p^\pr \phi_{n,l,m_l}^\ast({\bf p}^\pr)
 V({\bf p}^\pr,s_3,s_4;{\bf p},s_1,s_2)
\phi_{n,l,m_l}({\bf p})\;,
\label{eq:dirac}\eea
where
\bea
V({\bf p}^\pr,s_3,s_4;{\bf p},s_1,s_2)&=&
 \sqrt{
 \frac{J(p)}{16 \pi^3}\frac{J(p^\pr)}{16 \pi^3}} \left(
 V_{_{exchange}}+V_{_{annihil}}\right)
 - \sqrt{
 \frac{
 \left(\frac{1}{2 m}\right)}{16 \pi^3}\frac{\left(\frac{1}{2 m}\right)}{16 \pi^3}
 } 
 \left(V_{_{C}}\right)
\;.\label{eq:fermi}
\eea
The Dirac notation in Eq.~(\ref{eq:dirac})
will be used in the remainder of this paper.
See Eqs.~(\ref{eq:ruo}), (\ref{eq:saeko1}) and (\ref{eq:saeko2}) for $V_{_{C}}$,
$V_{_{exchange}}$ and $V_{_{annihil}}$ respectively. The interaction $V$ must
be diagonalized in the degenerate spin space following the standard rules 
of degenerate BSPT. Note that $V$ needs to be considered in second order
BSPT in this paper also.

The diagonalization of $V$ 
in the degenerate spin space
 follows shortly, but first
 recall from \S II  the range of $\lam$ that allows 
 the effects of low-energy (energy transfer below $\lam$)
   photon emission and absorption 
 to be  transferred to the effective interactions in the $|e {\overline e}\rangle$
 sector alone, 
 and at the same time  does not
remove the nonperturbative bound-state physics of interest:
\bea
&&\frac{\left|M_N^2-(2 m)^2\right|}{{\cal P}^+} \ll \frac{\lam^2}{{\cal P}^+} \ll q^{-}_{_{photon}}
\;.
\label{eq:theone}
\eea
After the solutions of ${\cal H}_o$ are known the $\alpha$-scaling in all BSPT matrix elements
is known and the bounds in Eq.~(\ref{eq:theone}) become
\bea
&&m^2 \alpha^2 \ll \lam^2 \ll m^2 \alpha
\;.\label{eq:132p}
\eea
This is satisfied  under the following limit:
\bea
\lam^2\;&\longrightarrow&{\rm a\;fixed\;number}~,\label{eq:limit2}\\
\frac{ m^2\alpha^2}{\lam^2}  &\longrightarrow& 0~,\\
 \frac{m^2\alpha }{\lam^2} &\longrightarrow& \infty
\;.\label{eq:limit3}
\eea
Given the nonrelativistic limit:
\bea
\alpha &\longrightarrow& 0~,\\
\frac{m^2}{\lam^2}&\longrightarrow&\infty
\;,
\eea
this implies 
\bea
&&\frac{m^2}{\lam^2}\propto
 \alpha^{-\frac{k}{2}} 
\;,\label{eq:2000}
\eea
where
\bea
&&2 < k <4~~.
\eea
Note that this ``window of opportunity" is available to us because one, we have introduced
an adjustable similarity hamiltonian energy scale, $\frac{\lam^2}{{\cal P}^+}$, 
into the theory, and two, $QED$ is 
a theory with two dynamical energy scales, $\frac{m^2 \alpha^2}{{\cal P}^+}$ and 
$\frac{m^2 \alpha}{{\cal P}^+}$, a fact known for a long time, and the reason that $QED$ 
calculations have been so successful over the years.

Given the above limit (Eqs. (\ref{eq:limit2})-(\ref{eq:limit3})),
\bea
&\bullet&\theta\left(\lam^2-4|{\bf p}^2-{{\bf p}^\pr}^2|\right)
 , \theta\left(|\Delta_1|-\lam^2\right) ,
\theta\left(|\Delta_2|-\lam^2\right)\longrightarrow 1
\label{eq:w}\\
&\bullet&\theta\left(4({\bf p}^2+m^2)-\lam^2\right) ,
\theta\left(4 ({{\bf p}^\pr}^2+m^2)-\lam^2\right) \longrightarrow 1
\label{eq:ww}\\
&\bullet&
  \theta\left(\lam^2-|\Delta_1|\right) ,
\theta\left(\lam^2-|\Delta_2|\right)\longrightarrow 0\\
&\bullet&\theta\left(\lam^2-4({\bf p}^2+m^2)\right) ,
\theta\left(\lam^2-4({{\bf p}^\pr}^2+m^2)\right) \longrightarrow 0
\;.\label{eq:1888}
\eea
Now we proceed with the diagonalization of $V$
 in the degenerate spin space (see Eqs.~(\ref{eq:dirac}) and (\ref{eq:fermi})).
 We will calculate all  corrections to order $\alpha^4$ that arise in the
spin splitting structure of the ground state spectrum of ${\cal H}_o$ 
via BSPT. 
First, we write $V$ more explicitly given the above limits
 in Eqs.~(\ref{eq:w})-(\ref{eq:1888}):
\bea
V({\bf p}^\pr, s_3,s_4; {\bf p}, s_1, s_2) &=&
\frac{1}{16 \pi^3} \frac{1}{2 m} \left(
1-\frac{{\bf p}^2+2 \kap_z^2+{{\bf p}^\pr}^2+2 {\kap_z^\pr}^2}
{4 m^2}+{\cal O}\left(\frac{{\bf p}^4}{m^4}
\right)
\right)\nonumber\\
&\times& \left(
-\frac{e^2 N_1}{DEN_4}-\frac{4 e^2}{(x-x^\pr)^2}\del_{s_1 s_3} \del_{s_2 s_4}
 +\frac{e^2 N_2}{DEN_5}+4 e^2 \del_{s_1 {\overline s}_2} \del_{s_3 {\overline s}_4}
\right)\nonumber\\
&-&\frac{1}{16 \pi^3}\frac{1}{2 m} V_{_{C}}
\;,
\eea
where
\bea
&\bullet& \frac{1}{DEN_4} \equiv \frac{\theta_{12}}{DEN_1}+
\frac{\theta_{21}}{DEN_2}\;\;,\;\theta_{12}\equiv
\theta\left(DEN_1-DEN_2\right)\\
&\bullet& \frac{1}{DEN_5}\equiv \frac{\theta\left({\cal M}_o^2-{{\cal M}_o^\pr}^2\right)}{{\cal M}_o^2}
+\frac{\theta\left({{\cal M}_o^\pr}^2-{{\cal M}_o}^2\right)}{{{\cal M}_o^\pr}^2}
\;.
\eea
Note that we have expanded out the Jacobian factors as given by Eq.~(\ref{eq:jac}).
Also, $DEN_1$ and $DEN_2$ are defined below Eq.~(\ref{eq:76}) and written in their expanded
version in Eqs.~(\ref{eq:ohno}) and (\ref{eq:ohno2}) respectively.
Finally, $N_1$ and $N_2$ are written below Eqs.~(\ref{eq:76}) and (\ref{eq:747}) respectively.

Now, since the
eigenstate wavefunctions of 
${\cal H}_o$  force 
${\bf p}$ to scale as ${\bf p} \sim m \alpha$, it is useful
to note the  $\alpha$-scaling 
of specific terms in $V$. Recalling  that 
we are always 
 assuming $\alpha \longrightarrow 0$
 (without which our matrix elements would 
 not have a well-defined $\alpha$ scaling), we see the following structure arising
\bea
V&=&V^{^{(0)}}+V^{^{(1)}} +V^{^{(2)}}+ \cdots\;,
\eea
where $V^{^{(S)}}$ scales as $V^{^{(S)}}\!\!\!\sim\alpha^S$.
Thus in first-order BSPT these respective terms contribute
\bea
\del^{^{(1)}}M_{N N^\pr}^2 =
\langle \phi_N|V^{^{(S)}}|\phi_{N^\pr} \rangle 
 \sim \alpha^{3 + S}
\;.\label{eq:80}
\eea
So, to be consistent to order $\alpha^4$ we need to look at all the matrix 
 elements 
 $V^{^{(S)}}$ with $S \leq 1$.\footnote{
For example,
 $ \frac{e^2 {\bf p}}
 {({\bf p}-{\bf p}^\pr)^2} 
 \sim\frac{\alpha^2}{\alpha^2} 
 \Rightarrow S = 0\;$.}
 
 Before proceeding to write out these expressions for $V^{^{(S)}}$,
  we note the following facts:
 
 $\bullet$ In this work we will only calculate  the  spin {\em splittings},
  so any constants along the diagonal in spin space do not contribute.
  
  $\bullet$ Since we are only working to order $\alpha^4$, obviously any splittings that can be shown
  to contribute at order $\alpha^{4+k}$ with $k > 0$ need not be calculated.
 
$\bullet$ Symmetries of the integrand can be used to simplify expressions
immensely.
 
 One final discussion that we must have,
  before we  write out these expressions for $V^{^{(S)}}$, is how we are going
  to deal with $DEN_4$ and 
  $DEN_5$ defined above.\footnote{Actually the $DEN_5$ term is handled with analogous 
  techniques as the $DEN_4$ term, and has even smaller corrections than those of $DEN_4$.
   Thus, we will just discuss the $DEN_4$ term in what follows and here state the result
   for the $DEN_5$ term: Take $DEN_5 \longrightarrow 4 m^2$; 
   the corrections to this start shifting the bound state mass
   at order $\alpha^6$.}
  These denominators are dealt with by noting
  the following formulae:
  \bea
\frac{\theta(a-b)}{a}+\frac{\theta(b-a)}{b}&=&
\frac{1}{2} \frac{\theta(a-b)+\theta(b-a)}{a}+
\frac{1}{2} \frac{\theta(a-b)+\theta(b-a)}{b} \nonumber\\
&+&\frac{1}{2} \frac{\theta(a-b)-\theta(b-a)}{a}-
\frac{1}{2} \frac{\theta(a-b)-\theta(b-a)}{b}\nonumber\\
&=&\frac{1}{2}\left( \frac{1}{a}+\frac{1}{b}\right)
+\frac{1}{2} \left(\theta(a-b)-\theta(b-a)\right)\left(
\frac{1}{a}-\frac{1}{b}\right)  \nonumber \\
&=&\frac{1}{2}\left( \frac{1}{a}+\frac{1}{b}\right)+\frac{1}{2} \frac{\left| a-b \right|}
{a-b} \left(
\frac{1}{a}-\frac{1}{b}\right)\nonumber\\
&=&\frac{1}{2}\left( \frac{1}{a}+\frac{1}{b}\right)-\frac{1}{2} \frac{\left| a-b \right|}
{a~b}
\;.\label{eq:seco}
  \eea
 To proceed it is useful to  note
 \bea
  DEN_1&=& ({\bf p}-{\bf p}^\pr)^2
 -\frac{(\kap_z-\kap_z^\pr) ({\bf p}^2-{{\bf p}^\pr}^2)}{m}
 +{\cal O}\left[\left(\frac{{\bf p}}{m}\right)^5 m^2\right]~,\\
 DEN_1&=&DEN_2-\frac{2 (\kap_z-\kap_z^\pr) ({\bf p}^2-{{\bf p}^\pr}^2)}{m}
 +{\cal O}\left[\left(\frac{{\bf p}}{m}\right)^5 m^2\right]~,\\
 \frac{1}{2}\left(\frac{1}{DEN_1}+\frac{1}{DEN_2}
 \right)&=&\frac{1}{\pp}+\frac{(\kap_z-\kap_z^\pr)^2({\bf p}^2-{{\bf p}^\pr}^2)^2}
 {m^2 ({\bf p}-{\bf p}^\pr)^6}
 +{\cal O}\left[\left(\frac{{\bf p}}{m}\right)^2\frac{1}{ m^2}\right]
 \;.\label{eq:rrr}
 \eea
 Especially note that this last equation scales as: $\frac{1}{\alpha^2}
 +1+\alpha^2+\cdots$, i.e. the corrections start at order $1$ (not order $\frac{1}{\alpha}$);
 this implies that only the $\frac{1}{\pp}$ term 
 of Eq.~(\ref{eq:rrr}) contributes to the spin splittings to 
 order $\alpha^4$. But we still have to discuss the second term that arises in Eq.~(\ref{eq:seco}).
 This term is given by 
 \bea
 \left.\frac{1}{DEN_4}\right|_{_{second~term}} &=&
 -\frac{1}{2} \frac{\left| DEN_1-DEN_2 \right|}
{DEN_1 DEN_2}=-\frac{1}{2}\frac{\left|\frac{2 (\kap_z-\kap_z^\pr) ({\bf p}^2-{{\bf p}^\pr}^2)}{m}\right|}
{\left({\bf p}-{\bf p}^\pr\right)^4}+{\cal O}\left(\alpha^{0}\right)
 \;.\label{eq:152}
 \eea
 Including $N_1$, this starts out as an ${\cal O}(\alpha^3)$ spin conserving
 contribution. The next order contribution is ${\cal O}(\alpha^4)$ with spin structure,
 but is odd under ${\bf p} \leftrightarrow
 {\bf p}^\pr$, and thus integrates to zero. However, the ${\cal O}(\alpha^3)$
 spin conserving term 
  appears to lead to an order $\alpha^4$ shift to the spin splittings
 in second order BSPT when the cross terms with $V^{^{(0)}}$ of Eq.~(\ref{eq:rzero})
 are considered;\footnote{$V^{^{(0)}}$ of Eq.~(\ref{eq:rzero}) comes from the first
 term of Eq.~(\ref{eq:rrr}) combined with the complete
 next to leading order term in
 $N_1$.} however,
 these cross term contributions add to  zero due to the facts that
 the ${\cal O}(\alpha^3)$ term including Eq.~(\ref{eq:152}) conserves spin and
 the ${\cal O}(\alpha^3)$ term including Eq.~(\ref{eq:152}) is even while the term from
 Eq.~(\ref{eq:rzero}) is odd under ${\bf p} \leftrightarrow
 {\bf p}^\pr$. 
 
 To summarize the  preceding discussion
 of $DEN_4$ and $DEN_5$, we can say that through order $\alpha^4$, for the spin splittings
 of  positronium,
 there are no relativistic corrections to the following replacements: 
 \bea
 &&DEN_4\longrightarrow \pp~~{\rm and}~~DEN_5\longrightarrow 4 m^2
 \;.
 \eea
  This is valid for the ground and excited states, but in what
  follows we specialize to the ground state for simplicity.
 
 Given this general conclusion about $DEN_4$ and $DEN_5$, 
  we list the pieces of $V$ that contribute to positronium's ground state
  spin splittings
 through order $\alpha^4$.
  Explicitly, as far as the $\alpha$-scaling goes, we need to consider
 \bea
 V^{^{(0)}}({\bf p}^\pr s_3 s_4;
 {\bf p} s_1 s_2)&=& \frac{- c_{_{ex}} e^2}{4  \pi^3 ({\bf p}-{\bf p}^\pr)^2}
  { v}^{(0)}({\bf p}^\pr s_3 s_4;
 {\bf p} s_1 s_2)
 \;,\label{eq:rzero}
 \eea
 where
 \bea
 { v}^{(0)}({\bf p}^\pr s_3 s_4;
 {\bf p} s_1 s_2)&\equiv&\left(\del_{s_1 {\overline s}_3} \del_{s_2 s_4} f_1({\bf p}^\pr s_3 s_4;
 {\bf p} s_1 s_2)+
 \del_{s_1 s_3} \del_{s_2 {\overline s}_4} f_2({\bf p}^\pr s_3 s_4;
 {\bf p} s_1 s_2)\right)\label{eq:132}~,\\
 f_1({\bf p}^\pr s_3 s_4;
 {\bf p} s_1 s_2) &\equiv& 
 s_1 (\kap_y-\kap_y^\pr)-i (\kap_x-\kap_x^\pr) ~,\\
 f_2({\bf p}^\pr s_3 s_4;
 {\bf p} s_1 s_2) &\equiv& s_4 (\kap_y-\kap_y^\pr)+i (\kap_x-\kap_x^\pr) 
 \;.
 \eea
 Recall that $s_i = \pm 1~(i=1,2,3,4)$ only. The only other interaction we need to consider is
 \bea
 V^{^{(1)}}({\bf p}^\pr s_3 s_4;
 {\bf p} s_1 s_2)&=& \frac{e^2}{4 m \pi^3}\left(c_{_{an}} \del_{s_1 s_2} \del_{s_1 s_4}
  \del_{s_3 s_4}+ c_{_{ex}} \del_{s_2 {\overline s}_4} \del_{s_2 {\overline s}_1}
  \del_{s_3 {\overline s}_1}+\right.\nonumber\\
 &&~~~~+ \left.\left(c_{_{an}} \frac{1}{2} - c_{_{ex}}
 \frac{(\kap-\kap^\pr)^2}{({\bf p}-{\bf p}^\pr)^2}\right)
  \del_{s_1 {\overline s}_2} \del_{s_3 {\overline s}_4}\right)\;.
 \eea
 The constants $c_{_{ex}}$ and $c_{_{an}}$
  were introduced only to distinguish the terms that arise from the `exchange' and 
 `annihilation' channels respectively, and $c_{_{ex}} = c_{_{an}} = 1$ will be used in the
 remainder of this work. 
 
 Two simplifications were made in 
 deriving  $V^{^{(1)}}$.  
   First,  
 we did not include terms that are a constant along the diagonal in
 spin space, because these do not contribute to the spin {\em splittings}
 to order $\alpha^4$. Second,
  we noted that
 terms of the following type integrate to zero:
 \bea
&& \langle \phi_{_{1,0,0,s_3,s_4}}|\frac{e^2 (\kap_x \kap_y^\pr,\kap_z \kap_x,\kap \times 
\kap^\pr)}{({\bf p}-
{\bf p}^\pr)^2}|\phi_{_{1,0,0,s_1,s_2}} \rangle=(0,0,0)
 \;.
 \eea
 and thus were not included in the definition of
 $V^{^{(1)}}$.

The ground state spin splitting to order $\alpha^4$ contains
 contributions from $V^{^{(1)}}$ in
first order BSPT ($V^{^{(0)}}$ vanishes in first order BSPT) and $V^{^{(0)}}$ 
in second order BSPT.
We  begin with the first order BSPT calculation. These results are shown in Figure~2.
Then we perform the second order BSPT calculation. The combined results of first and 
second order BSPT are shown in Figure~3.
\newpage
 \noindent
 \underline{First Order BSPT}:
 
The lowest order wavefunctions are given near the end of \S III.B.1 (see Appendix~B
for the hyperspherical harmonics). $V^{^{(1)}}$ in first order BSPT contributes
the following to positronium's ground state mass squared: 
 \bea
\del M_{_{1}}^2&\equiv&
 \del^{^{(1)}}M^2(s_3,s_4;s_1,s_2)\nonumber\\
 &=& N \int d^3 p~ d^3 p^\pr \frac{1}{(e_1^2+{\bf p}^2)^2} 
 \frac{1}{(e_1^2+{{\bf p}^\pr}^2)^2} V^{^{(1)}}({\bf p}^\pr,s_3,s_4;{\bf p},s_1,s_2)
 \;,
 \label{eq:178}
 \eea
 where
 \bea
 N&=&\frac{8 e_1^5}{\pi^2}\;\;{\rm and}\;\;e_1 = \left.\frac{m \alpha}{2 n}\right|_{n=1}
 \;.
 \eea
 Using the rotational symmetry of the integrand we can make the substitution
 \bea
 \frac{(\kap-\kap^\pr)^2}{\pp}&\longrightarrow&\frac{\frac{2}{3}\left(
 (\kap_x-\kap_x^\pr)^2+(\kap_y-\kap_y^\pr)^2+(\kap_z-\kap_z^\pr)^2\right)}{\pp}
 =\frac{2}{3}
 \;.
 \eea 
 After this,  the remaining integrals are trivial and the splittings that arise from diagonalization
 of the $\easy$ matrix in spin space are:
 \bea
 \left\langle 1\left |\easy\right| 1 \right\rangle&=&- m^2 \alpha^4~,\\
  \left\langle 2\left |\easy\right| 2 \right\rangle&=& \frac{2}{3} m^2 \alpha^4~,\\
  \left\langle 3\left |\easy\right|3 \right\rangle&=& m^2 \alpha^4~,\\
  \left\langle 4\left |\easy\right| 4 \right\rangle&=& m^2 \alpha^4
 \;,
 \eea
 where
 \bea
&&\left\{ |1\rangle\equiv\frac{|+-\rangle-|-+\rangle}{\sqrt{2}}\;,\;
|2\rangle\equiv\frac{|+-\rangle+|-+\rangle}{\sqrt{2}}\;,\;
|3\rangle\equiv|--\rangle\;,\;
|4\rangle\equiv|++\rangle\right\}
 \;.\label{eq:187}
 \eea
Figure~2 shows these results, which taken alone do not produce the degeneracies required by
rotational invariance. 
 
  \noindent
 \underline{Second Order BSPT}:
 
 $V^{^{(0)}}$ gives rise to the following contribution 
 to positronium's ground state mass squared
 in second order BSPT:
 \bea
 \hard&\equiv&
\del^{^{(2)}}M^2(s_3,s_4;s_1,s_2)\nonumber\\
&=&\sum_{s_e,s_{{\overline e}}} \sum_{\mu \neq (1,0,0)}
\frac{\langle \phi_{1,0,0,s_3,s_4} | V^{^{(0)}} | \phi_{\mu,s_e,s_{{\overline e}}}
\rangle \langle \phi_{\mu,s_e,s_{{\overline e}}}|  V^{^{(0)}}| \phi_{1,0,0,s_1,s_2} \rangle}
{{\cal M}_1^2-{\cal M}_n^2}
  \;.
  \label{eq:happyman}
  \eea
  Recall that $\mu = (n,l,m_l)$, the usual principal and angular momentum quantum 
  numbers of nonrelativistic positronium.
  The calculation of $\hard$ is tedious, but can be done analytically.
  This calculation is performed in
  Appendix~C. The result is (see Eq.~(\ref{eq:whatthehell}) in Appendix~C): 
  \bea
 \hard &=& - \frac{m^2 \alpha^4}{24} (3 g_1+ g_2) 
 \;,
 \eea
where $g_1$ and $g_2$ are given in Appendix~C in Eqs.~(\ref{eq:g1}) and (\ref{eq:g2}) respectively.

 Now we  combine the $\easy$ and $\hard$ matrices and diagonalize the result. The combined matrix is
 given by:
 \bea
 \frac{\easy+\hard}{2 m^2 \alpha^4}&=& \frac{1}{2} \del_{s_1 s_2} \del_{s_1 s_4} \del_{s_3 s_4} 
 -\frac{1}{12} \del_{s_1 {\overline s}_2} \del_{s_3 {\overline s}_4}+\frac{1}{2} 
 \del_{s_2 {\overline s}_4} \del_{s_1 {\overline s}_2} \del_{s_1
  {\overline s}_3}\nonumber\\
  &&~~~~~-\frac{1}{48} (3 g_1+g_2)\;.
 \eea
 The eigenvalues are:
  \bea
 \left\langle 1 \left|\easy+\hard\right| 1 \right\rangle&=&- \frac{5}{3} m^2 \alpha^4\label{eq:1776}~,\\
 \left\langle 2 \left|\easy+\hard\right| 2 \right\rangle&=& \frac{2}{3} m^2 \alpha^4~,\\
  \left\langle 3 \left|\easy+\hard\right| 3 \right\rangle&=&  \frac{2}{3} m^2 \alpha^4~,\\
 \left\langle 4 \left|\easy+\hard\right| 4 \right\rangle&=&  \frac{2}{3} m^2 \alpha^4
 \;,\label{eq:1493}
 \eea
 and the corresponding eigenvectors are:
 \bea
&&\left\{ |1\rangle\equiv\frac{|+-\rangle-|-+\rangle}{\sqrt{2}}\;,\;
|2\rangle\equiv\frac{|+-\rangle+|-+\rangle}{\sqrt{2}}\;,\;
|3\rangle\equiv|--\rangle\;,\;
|4\rangle\equiv|++\rangle\right\}
 \;.
 \eea
  Figure~3 displays these results. 
 These results lead to the well-known result, $\frac{7}{6} \alpha^2 Ryd$, as
 detailed in Appendix~D.
We see rotational invariance in the degeneracies of the 
ground state $n=1$
levels
exactly maintained at order $\alpha^4$.

%%%%%%%%%%%%%%%%%%%%%%%%%%%%%%%%%%%%%%%%%%%%%%%%%%%%%%%%%%%%%%%%%%%%%%%%%%%%%
%%%%%%%%%%%%
%%%%%%%%%%%%
%%%%%%%%%%%%                
%%%%%%%%%%%%
%%%%%%%%%%%%%%%%%%%%%%%%%%%%%%%%%%%%%%%%%%%%%%%%%%%%%%%%%%%%%%%%%%%%%%%%%%%%%%
\newpage
\noindent
\centerline{{\bf IV. REMAINING CORRECTIONS TO THE POSITRONIUM SPECTRUM}}
\vskip.2in

We have shown how one obtains spin {\em splittings} at 
${\cal O}(\alpha^4)$, starting with a light-front Hamiltonian renormalized
through ${\cal O}(\alpha)$.  Even though the 
spin splittings are correct through ${\cal O}(\alpha^4)$,
 there are corrections starting at ${\cal
O}(\alpha^3)$ in the actual position of the levels computed using $H_{_{R}}$
renormalized to
${\cal O}(\alpha)$ and $m^2 \alpha^2 \ll \lam^2 \ll m^2 \alpha$.   
We will outline these corrections
 and argue that higher order counterterms, obtained from renormalizing
the Hamiltonian through ${\cal O}(\alpha^2)$ and beyond, fix any errors
appearing in the physical spectrum computed with 
$H_{_{R}}$
renormalized to
${\cal O}(\alpha)$.

The dressing of electrons by two photon loops in all allowed
time-orderings shift the electron self-energy by ${\cal O}(\alpha^2
\lambda^2 \log^2(1/\epsilon))$.  In an analytic
calculation this contribution can be cancelled by a one-body \linebreak
counterterm that
is exactly what one expects from higher order renormalization   when the Hamiltonian
is computed to ${\cal O}(\alpha^2)$. The infrared divergences 
must cancel leaving a free electron energy $\frac{{p^\perp}^2+m^2}{p^+}$.

The leading correction to the spectrum we have computed in this paper may be
${\cal O}(\alpha^2 \lambda^2 \log^2(1/\epsilon))$, coming from 
the above ${\cal O}(\alpha^2)$
correction to the electron self-energy.  However, with $H_{_{R}}$ 
renormalized  through ${\cal O}(\alpha)$, there are
indications that all the infrared divergences arising in BSPT cancel 
due to the fact that large wavelength photons decouple from  neutral
positronium states \cite{perrybrazil}, leading only to residual
long-range two-body potentials that may still correct the spectrum.
Assuming this holds at higher orders too,
the leading correction from higher order self-energy shifts
is ${\cal O}(\alpha^2 \lambda^2 )$.
  With the cutoffs
used above the corrections to the electron self-energy can depend on the
longitudinal momentum fraction of the electron, while if  cutoffs 
involving the momenta of the {\em connected} pieces of the matrix elements 
 are used, the relevant counterterm is a constant.  

From Eq.~(\ref{eq:2000}) in the paper, the constraint
placed on $\lam^2$ is
\bea
&&\lam^2\propto
m^2 \alpha^{\frac{k}{2}} \:,
\eea
where $2<k<4$. So the leading order shift to the spectrum arising
from higher order self-energy corrections is
\bea
&&\del M^2_{c1} \sim m^2 \alpha^2\alpha^{\frac{k}{2}}\;.
\eea

Another correction to the spectrum from the ${\cal O}(\alpha)$ Hamiltonian
arises from the interaction containing the leading order term of $N_1$ and
the term from Eq.~(\ref{eq:152}). This shifts the spectrum an amount
\bea
&&\del M^2_{c2} \sim m^2 \alpha^3\;.
\eea

So, after considering the possible
corrections, we see that our choice
of ${\cal H}_o$ has produced
eigenvalues that are correct through ${\cal O}(m^2 \alpha^2)$.
That is, the nonperturbative physics was correctly chosen and is not changing.
However, fourth order perturbative effective interactions are
 required to obtain ${\cal O}(m^2\alpha^4)$ precision
 for the actual position of the levels---the spin splittings are correct through
 order
 $\alpha^4$ as detailed in the paper. 

To be complete, we must also consider coupling and photon mass
renormalization. 
However, there are no further corrections
to the photon mass if $\lambda < 2m$ because pair production is no
longer possible. This also affects coupling renormalization.
 If $\lambda < 2 m$, no electron
loop appears to modify the electron-photon vertex.  Since this contribution is
entirely responsible for renormalization of the marginal part of this operator
({\it i.e.}, the running coupling), we do not expect the charge to renormalize. 
There will be modifications to the irrelevant part of this vertex, but they
should not contribute to leading order.  

There are additional counterterms that are not typically encountered in a
lagrangian calculation.  For example, there are tree level counterterms  that
modify the electron-photon and positron-photon coupling at ${\cal O}(e^3)$. 
These counterterms should produce cancellations with photon exchange
contributions to the mass, but they must be present for this to occur.  Such
counterterms may be necessary for all the infrared divergences to cancel.

We offer no full solution to these problems other than renormalizing the
Hamiltonian to higher orders using a similarity transformation and coupling
coherence, as outlined and illustrated in earlier sections.

%%%%%%%%%%%%%%%%%%%%%%%%%%%%%%%%%%%%%%%%%%%%%%%%%%%%%%%%%%%%%%%%%%%%%%%%%%%%
%%%%%%%%%%%%
%%%%%%%%%%%%
%%%%%%%%%%%%                here's section ?????
%%%%%%%%%%%%
%%%%%%%%%%%%%%%%%%%%%%%%%%%%%%%%%%%%%%%%%%%%%%%%%%%%%%%%%%%%%%%%%%%%%%%%%%%%%%
%%%%%%%%%%%%%%%%%%%%%%%%%%%%%%%%%%%%%%%%%%%%%%%%%%%%%%%%%%%%%%%%%%%%%%%%%%%%%
%%%%%%%%%%%%
%%%%%%%%%%%%
%%%%%%%%%%%%                here's summary of results and final remarks
%%%%%%%%%%%%
%%%%%%%%%%%%%%%%%%%%%%%%%%%%%%%%%%%%%%%%%%%%%%%%%%%%%%%%%%%%%%%%%%%%%%%%%%%%%%
\vskip.2in
\noindent
\centerline{{\bf V. SUMMARY AND DISCUSSION}}
\vskip.2in

 If $H_{_{R}}$ is derived  approximately to a finite order
in $\alpha$, to what order in $\alpha$ is the nonperturbative
spectrum correct? The precise 
connection
between the approximate renormalized Hamiltonian and the physical
spectrum is only qualitatively understood at present. The results of this paper
 make it clear that in light-front QED to obtain positronium's spin
 structure through order
$\alpha^5$, the renormalized Hamiltonian must be  derived either through at least second order
in $\alpha$ or else via a nonperturbative similarity transformation which
uses Coulomb states instead of free states in the  
 perturbative expansion for the effective
interactions. 

We have calculated the order $\alpha^4$ 
spin {\em splittings} for the $n=1$ levels.  This restriction leads to a great 
simplification because constants along the diagonal in spin space do not contribute to the {splittings}.
Some of the terms that are not calculated because they cancel in the {difference}
 are quite complicated
and a future analytical 
calculation including these terms would be quite complex,
 involving complicated sums over
9-J symbols for example.
  Even though we do not calculate these $\alpha^4$ ``radial" shifts in
  this paper, the 
fact that our spin splittings are correct
 and can be computed
 analytically from first principles gives us much 
hope that the procedure outlined in this work is well-defined and at least for $QED_{_{3+1}}$
leads to consistent bound state calculations. 
It becomes very interesting to think about the Lamb shift.
%%%%%%%%%%%%%%%%%%%%%%%%%%%%%%%%%%%%%%%%%%%%%%%%%%%%%%%%%%%%%%%%%%%%%%%%%%%%%
%%%%%%%%%%%%
%%%%%%%%%%%%
%%%%%%%%%%%%               
%%%%%%%%%%%%
%%%%%%%%%%%%%%%%%%%%%%%%%%%%%%%%%%%%%%%%%%%%%%%%%%%%%%%%%%%%%%%%%%%%%%%%%%%%%%
%%%%%%%%%%%%%%%%%%%%%%%%%%%%%%%%%%%%%%%%%%%%%%%%%%%%%%%%%%%%%%%%%%%%%%%%%%%%%
%%%%%%%%%%%%
%%%%%%%%%%%%
%%%%%%%%%%%%                here's the nice people
%%%%%%%%%%%%
%%%%%%%%%%%%%%%%%%%%%%%%%%%%%%%%%%%%%%%%%%%%%%%%%%%%%%%%%%%%%%%%%%%%%%%%%%%%
\vskip.2in
\noindent
\centerline{{\bf  ACKNOWLEDGMENTS}}
\vskip.2in
The authors wish to acknowledge useful discussions with Brent H. Allen, 
Martina M. Brisudov\'{a}, James B. White and Kenneth G. Wilson. In particular,
we would like to thank Brent H. Allen for the use of his
matrix element rules \cite{brent}.
Research reported in this paper has been supported  by 
the National Science Foundation under grant PHY-9409042 and
the Maria Sk{\l}odowska-Curie Foundation
under grant MEN/NSF-94-190.
%%%%%%%%%%%%%%%%%%%%%%%%%%%%%%%%%%%%%%%%%%%%%%%%%%%%%%%%%%%%%%%%%%%%%%%%%%%%%
%%%%%%%%%%%%
%%%%%%%%%%%%
%%%%%%%%%%%%                here's the appendices
%%%%%%%%%%%%
%%%%%%%%%%%%%%%%%%%%%%%%%%%%%%%%%%%%%%%%%%%%%%%%%%%%%%%%%%%%%%%%%%%%%%%%%%%%%%
\newpage
\noindent
\centerline{{\bf  APPENDIX A: LIGHT FRONT CONVENTIONS}}
\vskip.2in

  In this appendix we present our light-front conventions.
  In this initial paragraph, let
$A$ and $B$ be  arbitrary 4-vectors. An ``ET" label implies an equal-time vector; the 
absence of a label implies a   light-front vector.
Two vectors by definition are related by
\beaa
&\bullet&A^{\pm} = A^0_{ET} \pm A^3_{ET}\\
&\bullet&A^{i} = A^{i}_{ET} ,\; i=1,2\\
&\bullet&A^\mu=(A^+,A^-,A^i)\\
&\bullet&  A^{\mu}_{ET}=(A^0_{ET},A^3_{ET},A^i).
\eeaa
Then scalars are required to agree by adjusting the metric tensor so
that this is so. This fixes the light-front metric tensor:
\beaa
&& A^\mu B_\mu= g_{\mu \nu} A^\mu B^\nu =
A^{\mu}_{ET} B_{\mu}^{ET}= g_{\mu \nu}^{ET} A^{\mu}_{ET} B^{\nu}_{ET} =
A^0_{ET} B^0_{ET} -A^3_{ET} B^3_{ET}-A^{i}  B^{i}~.\\
&\Longrightarrow& 1=2 g_{+-}=2g_{-+}=
-g_{11}=-g_{22}\;.
\eeaa
 Another 
relevant scalar is of course
\beaa
&&\; g_{\mu\nu}g^{\mu \nu}=g_{\mu\nu}^{ET}g^{\mu \nu}_{ET}=4\;.\\
 &\Longrightarrow& 1=\frac{g^{+-}}{2}=\frac{g^{-+}}{2}= 
-g^{11}=-g^{22}\;.
\eeaa
The components of the light-front metric tensor
not mentioned are zero. 
Thus, there are factors of two in places like
\beaa
d^4x=\frac{1}{2}dx^+ dx^- d^2x^{\perp}\;,\; A_{-}=g_{-+}A^{+}=\frac{1}{2}
(A^0+A^3)=\frac{1}{2}(A_0-A_3)\;,\;etc.
\eeaa

Conventionally, $x^+$ is chosen to be the light-front time coordinate. This fixes $x^-$ to
be the light-front longitudinal space coordinate. Also, from the $p\cdot x$ scalar:
\beaa
p_\mu x^\mu&=& g_{+-} p^- x^{+} + g_{-+} p^+ x^{-}-p^i x^i=
\frac{1}{2} p^- x^{+} + \frac{1}{2} p^+ x^{-}-p^i x^i
~,\eeaa
we see that $p^-$ is fixed to be 
the light-front energy coordinate, and $p^+$ is fixed to be the light-front
longitudinal momentum coordinate.

 All constituents
in the forward light-cone in light-front coordinates have
$p^+ \geq 0$. This
can be seen from the following relation (which also shows why
particle lines in hamiltonian diagrams are all on mass shell):
\beaa
&&\int\frac{d^4p}{(2 \pi)^4}2 \pi \del (p^2-m^2) \theta(p^0)
f(p)=\int\frac{d^3p}{16\pi^3p_0}f(p){\Biggr{|}}_{p_0=\sqrt{{\vec{p}}^2+m^2}}\\
&=&\;\int\frac{d^2p^\perp dp^+ \theta(p^+)}{16\pi^3p^+}f(p)
{\Biggr{|}}_{p^-=\frac{{p^\perp}^2+m^2}{p^+}}\equiv \int_{p}f(p)
{\Biggr{|}}_{p^-=\frac{{p^\perp}^2+m^2}{p^+}}\;~~.
\eeaa
 Especially note this last definition of $\int_p$~, which is a shorthand used
 in the paper. 
 
  In momentum space the field
operators are expanded as (at $x^+ = 0$)
\beaa
A^i(x)&=&\sum_{s = \pm 1} \int_{q} (\eps^i_s a_s(q)
e^{-iq \cdot x}+h.c.)\;,\\
\xi(x)&=&\sum_{s = \pm 1}\chi_s \int_p \sqrt{p^+}(b_s(p)
e^{-ip \cdot x}+d_{{\overline s}}^{\dagger}(p)e^{+ip\cdot x})\;,\\
{\rm with}
&&\eps_1^i=\frac{-1}{\sqrt{2}}\left(\delta_{i,1} + i ~\delta_{i,2}\right)
\;,\;\eps_{-1}^i=\frac{1}{\sqrt{2}}\left(
\delta_{i,1} - i~ \delta_{i,2}\right)\;,\\
&&\chi_{_{1}}=
\left(\begin{array}{c}
1\\
0
\end{array}\right)\;,\;\chi_{_{\overline 1}}=
\left(\begin{array}{c}
0\\
1
\end{array}\right)
\;.
\eeaa
The fermion  helicity can only take
on the  values
$ \pm 1/2$, however we define
$h_3=  s/2$; therefore, ``s" can only take
on the values
$\pm 1$. Note that ${\overline s} \equiv -s$. Examples of the
commutation (anti-commutation) relations and free Fock states are 
\beaa
&&[a_\lambda(q),a_{\lambda^\prime}^{\dagger}(q^\prime)]=16\pi^3q^+
\del^3(q-q^\prime)\del_{\lambda \lambda^\prime}\;,\;(\;\del^3(p)\equiv
\del^2(p^\perp)\del(p^+)\;)\;,\\
&&\{b_s(p),b_{s^\prime}^{\dagger}(p^\prime)\}=
\{d_s(p),d_{s^\prime}^{\dagger}(p^\prime)\}=
16\pi^3p^+
\del^3(p-p^\prime)\del_{s s^\prime}\;,\\
&&\langle p_1 s_1 | p_2 s_2 \rangle = 16\pi^3p_1^+
\del^3(p_1-p_2)\del_{s_1 s_2}\;,\;|p_1 s_1\rangle=
b_{s_1}^{\dagger}(p_1)|0\rangle\;,\;etc.
\eeaa
The inverse longitudinal derivative can be defined as follows (we define it
by putting  the momentum representations of the field operators
in the Hamiltonian, multiplying
the terms out explicitly, and then replacing the inverse derivative by
appropriate factors of longitudinal momentum [$\times \pm i$] but
nevertheless the following could be used too):
\beaa
&&\Biggl{(} \frac{1}{\partial^+} \Biggr{)}f(x^-)=\frac{1}{4}
\int_{- \infty}^{+ \infty}dy^{-}\eps(x^{-} - y^{-})f(y^-)\;,\\
 &&\partial^+=2\partial_-=2\frac{\partial}{\partial x^-}\;,\\
&&\partial_- \eps(x^- - y^-)=2 \del(x^- - y^-)\;,\\
&&\eps(x)=\theta(x)-\theta(-x)\;.
\eeaa
Notice that this is non-local in the longitudinal direction.
%%%%%%%%%%%%%%%%%%%%%%%%%%%%%%%%%%%%%%%%%%%%%%%%%%%%%%%%%%%%%%%%%%%%%
%%%%%%%%%%%%%%%%%%%%%%%%%%%%%%%%%%%%%%%%%%%%%%%%%%%%%%%%%%%%%%%%%%%%%%%%%%
%%%%%%%%%%%%%%%%%%%%%%%%%%%%%%%%%%%%%%%%%%%%%%%%%%%%%%%%%%%%%%%%%%%%%%%%%%%%%%%
\vskip.2in
\noindent
\centerline{{\bf  APPENDIX B: HYPERSPHERICAL HARMONICS/FOCK COORDINATE CHANGE}}
 \vskip.2in
 
 In this appendix we will list some useful mathematical relations used in the paper.
 The conventions followed in this paper are given in Ref.~\cite{judd1}. These hyperspherical 
 harmonics are given by:
 \bea
 Y_\mu(\Omega)&\equiv&Y_{n,l,m}(\Omega)\equiv f_{n,l}(\omega) Y_{l,m}(\theta,\phi)
 \;,
 \eea
 where
 \bea
 0&\leq&|m|\leq l\leq n-1
 \;.
 \eea
 These quantum numbers are the standard ``hydrogen" quantum numbers.
 These 3D
  spherical harmonics, $Y_{l,m}(\theta,\phi)$, are given by~\cite{jackson}:
 \bea
 Y_{l,m}(\theta,\phi)&=&
 \sqrt{\frac{2 l+1}{4 \pi}\frac{(l-m)!}{(l+m)!}}
 \frac{(-1)^m}{2^l l!}
 (1-x^2)^{\frac{m}{2}}
 \frac{d^{l+m}}{dx^{l+m}}
 (x^2-1)^l
 e^{i m \phi}
 \;,
 \eea
 where $x=\cos\theta$.
 These other functions, $f_{n,l}(\omega)$, are given by:
 \bea
 f_{n,l}(\omega)&=&
 (-i)^l \sqrt{\frac{2 n(n-l-1)!}
 {\pi (n+l)!}}
 \sin^l\omega\frac{d^l}{(d\cos\omega)^l}C_{n-1}(\cos\omega)
 \;,
 \eea
 where $C_{n-1}(\cos\omega)$ are Gegenbauer polynomials. For example~\cite{judd1prime}:
 \bea
 C_0(y)&=&1\;,\;C_1(y)=2 y\;,\;C_2(y)=4 y^2-1\;,\;C_3(y)=8 y^3-4 y\nonumber~,\\
 C_4(y)&=&16 y^4-12 y^2+1 \;,\;C_5(y)=32 y^5-32 y^3+6 y
 \;.
 \eea
 The orthogonality and phase relations  of these above functions are:
 \bea
Y_{n,l,m}&=& (-1)^{l+m} Y_{n,l,-m}^\ast\;,\;Y_{l,m}=(-1)^m Y_{l,-m}^\ast\;,\;f_{n,l}=
(-1)^lf_{n,l}^\ast~, \\
d\Omega^{^{(4)}}&\equiv& d\Omega \equiv  d\Omega^{^{(3)}} d\omega \sin^2\omega~, \\
\int d\Omega Y_\mu^\ast Y_{\mu^\pr}&=&\del_{\mu \mu^\pr}\;,\;\int d\omega\sin^2\omega f_{n,l}^\ast
 f_{n^\pr,l}=\del_{n n^\pr}\;,\;\int d\Omega^{^{(3)}} Y_{l,m}^\ast Y_{l^\pr,m^\pr}=
 \del_{l l^\pr} \del_{m m^\pr}
\;.
\eea
The first few hyperspherical harmonics are:
\bea
Y_{1,0,0}&=&\frac{1}{\sqrt{2 \pi^2}}~~,~~Y_{2,0,0}=\frac{\sqrt{2} \cos \omega}{\pi}\nonumber\\
Y_{2,1,-1}&=&\frac{-i e^{-i\phi} \sin \omega \sin \theta}{\pi}~~,~~Y_{2,1,0}=\frac{-i \sqrt{2}
 \sin \omega\cos \theta}{\pi}\nonumber\\
Y_{2,1,1}&=&\frac{i e^{i\phi} \sin \omega \sin \theta}{\pi}\;.
\eea
For further harmonics we
refer the interested
reader to Appendix~2 of Judd's text~\cite{judd2}, where this is done quite nicely.

For the coordinate change in the Coulomb Schr{\"o}dinger equation 
(see  Eq.~(\ref{eq:C}) and the discussion that follows it),
for ${\cal B}_{_{N}} < 0$, we define
\bea
 m {\cal B}_{_{N}}&\equiv&-{e_n}^2~,\\
u&\equiv&(u_0,{\bf u}) ~,\\
 u_0&\equiv&\cos(\omega)\equiv\frac{e_n^2-{\bf p}^2}{e_n^2+{\bf p}^2}~,\\
 {\bf u}&\equiv&\frac{{\bf p}}{p}\sin(\omega) \equiv \sin(\omega)\left(
\sin(\theta) \cos(\phi) ,sin(\theta) \sin(\phi) , \cos(\theta)\right)\nonumber\\
&\equiv& \frac{2 e_n {\bf p}}{e_n^2+{\bf p}^2}
 \;.\eea
 Note that $u_0^2+{\bf u}^2=1$. Conversely this coordinate change gives:
 \bea
 {\bf p}&=&\frac{e_n}{1+u_0}{\bf u}~,\\
 e_n^2+{\bf p}^2&=&\frac{2 e_n^2}{1+u_0}
 \;.
 \eea
 We also have:
 \bea
 &&d\Omega_p= \sin^2\omega d\omega d\Omega_p^{^{(3)}}=\left(
\frac{2 e_n}{e_n^2+{\bf p}^2} \right)^3 d^3p~,\\
&&\del^3(p-p^\pr)=\frac{(2 e_n)^3}{(e_n^2+{\bf p}^2)^3}
 \del(\Omega_p-\Omega_{p^\pr})=\frac{(1+u_o)^3}{e_n^3} \del(\p-\ppr)~,\\
&& 0 \leq \omega\leq\pi\;,\;0\leq\theta\leq\pi\;,\;0\leq\phi\leq 2 \pi
 \;.
 \eea
 Finally, a most useful relation is:
 \bea
 |{\bf p}-{\bf p}^\pr|^2&=&\frac{(e_n^2+{\bf p}^2)(e_n^2+{{\bf p}^\pr}^2)}{4 e_n^2}|u-u^\pr|^2
 \;.
 \label{eq:1492}
 \eea
 This is useful because we can expand $|u-u^\pr|^2$ as follows:
 \bea
 \frac{1}{|u-u^\pr|^2}&=&
 \sum_\mu \frac{2 \pi^2}{n} Y_\mu(\Omega_p) Y_\mu^\ast(\Omega_{p^\pr})
 \;.
 \label{eq:1900}
 \eea
 This completes the discussion of the hyperspherical harmonic 
 mathematical relations used in this work; actually,
 as shown in the next appendix,
  the $\hard$ calculation requires some further formulae, which are given
 as they are needed.
%%%%%%%%%%%%%%%%%%%%%%%%%%%%%%%%%%%%%%%%%%%%%%%%%%%%%%%%%%%%%%%%%%%%%%%%%%
%%%%%%%%%%%%%%%%%%%%%%%%%%%%%%%%%%%%%%%%%%%%%%%%%%%%%%%%%%%%%%%%%%%%%%%%%%%%%%%
\vskip.2in
\noindent
\centerline{{\bf  APPENDIX C: THE CALCULATION OF} {\boldmath $\hard$} }
\vskip.2in

In this appendix we perform the following sum 
analytically:
\bea
\hard&=&\sum_{s_e,s_{{\overline e}}} \sum_{\mu \neq (1,0,0)}
\frac{\langle \phi_{1,0,0,s_3,s_4} | V^{^{(0)}} | \phi_{\mu,s_e,s_{{\overline e}}}
\rangle \langle \phi_{\mu,s_e,s_{{\overline e}}}| V^{^{(0)}}| \phi_{1,0,0,s_1,s_2} \rangle}
{{\cal M}_1^2-{\cal M}_n^2}
  \;.
  \eea 
  Recall that $\mu=(n,l,m_l)$, the usual principal and angular momentum quantum numbers of 
  nonrelativistic positronium.
 Also recall that the spin factored completely out of our lowest order Schr{\"o}dinger equation, so 
 to proceed the
 following notation is useful:
 \bea
 |\phi_{\mu,s_e,s_{{\overline e}}}
\rangle&=& |\phi_\mu\rangle \otimes |s_e s_{{\overline e}}\rangle
 \;,\\
 1&=& 
\sum_{s_e,s_{\overline e},\mu} |\phi_\mu\rangle\langle\phi_\mu|
  \otimes |s_e s_{{\overline e}}\rangle\langle s_e s_{\overline e}|\nonumber\\
  &=&\sum_{s_e,s_{\overline e}} \int d^3 p ~|{\bf p}\rangle\langle {\bf p} |
  \otimes |s_e s_{\overline e}\rangle\langle s_e s_{\overline e}|
  \;.
 \eea
 To proceed, define the following Green's function for  arbitrary $E$:
 \bea
 \frac{{G}_E}{4 m}&\equiv& \sum_\mu \frac{|\phi_\mu \rangle
 \langle \phi_\mu |}{E-{\cal M}_n^2}
 \;.
 \eea
 The factor $\frac{1}{4 m}$ will turn out to be useful. This Green's function 
  satisfies
 the familiar Coulomb Green's function equation:
 \bea
 \del^3({\bf p}-{\bf p}^\pr) &=&
 (\tilde{E}-\frac{{{\bf p}^\pr}^2}{m}) G_E({\bf p}^\pr,{\bf p})+
 \frac{\alpha}{2 \pi^2} \int d^3 p^{\pr \pr} \frac{G_E({\bf p}^{\pr \pr},{\bf p})}
 {({\bf p}^\pr-{\bf p}^{\pr \pr})^2}
 \;,
 \eea
 where
 \bea
 &&\langle {\bf p}^\pr | {G}_E |{\bf p}\rangle\equiv G_E({\bf p}^\pr,{\bf p})~,\\
 {\rm and}&&\;\tilde{E}\equiv\frac{E-4 m^2}{4 m}\;.
 \eea
  Hostler and Schwinger
   independently obtained the solution for this Coulomb Green's function 
 in 1964~\cite{schwinger}. We find Schwinger's form useful;
   the equation  he solves is exactly the above equation 
  with the following shifts in notation:
 \bea
 \left(Z e^2\right)_{Schwinger} &\longrightarrow& \alpha~,\\
 m_{Schwinger}&\longrightarrow& \frac{m}{2}~,\\
 E_{Schwinger}&\longrightarrow& \tilde{E}
 \;.
 \eea
His result is amended because the  sum we need has  $E =
{\cal M}_1^2$ and does not include $\mu = (1,0,0)$. 
This subtraction of the $\mu = (1,0,0)$ term amounts to the term ``$-\frac{1}{C}$" in $G_{III}$
below. The details of how this arises can be seen in Eqs.~(\ref{eq:billy1})-(\ref{eq:billy2}) below.
With
this amendment, Schwinger's result is:
\bea
G_{{\cal M}_1^2}^\pr({\bf p},{\bf p}^\pr)&\equiv&G_I+G_{II}+G_{III}~,\label{eq:G}\\
G_I&=&\frac{\del^3({\bf p}-{\bf p}^\pr)}{\tilde{E}-T}~,\\
G_{II}&=& -\frac{\alpha}{2 \pi^2} \frac{1}{\tilde{E}-T}\frac{1}{\pp}
\frac{1}{\tilde{E}-T^\pr}~,\\
G_{III}&=&-\frac{\alpha}{2 \pi^2} \frac{4 e_1^2}{\tilde{E}-T}
\left[\int_{0}^{1}\frac{ d \rho }{\rho} \left(
\frac{1}{4 e_1^2 \rho \pp+C (1-\rho)^2}-
\frac{1}{C}
\right)
\right]\frac{1}{\tilde{E}-T^\pr}~,
\eea
where
\bea
 T &=& \frac {{\bf p}^2}{m}\;,\;T^\pr =\frac{{{\bf p}^\pr}^2}{m}\;,\;e_1=\frac{m \alpha}{2}~,\\
C&=&(e_1^2+{\bf p}^2)(e_1^2+{{\bf p}^\pr}^2)~,\\
\tilde{E}&=&\frac{{\cal M}_1^2-4 m^2}{4 m}=-\frac{e_1^2}{m}
\;.
\eea
The prime on $G_{{\cal M}_1^2}^\pr$ denotes the fact that we
 have subtracted  the $\mu = (1,0,0)$ part of $G_E$ as required by the 
sum that we have in $\hard$. Note that this Green's function is symmetric under ${\bf p} 
\leftrightarrow {\bf p}^\pr$ and also $(p_x,p_x^\pr) \leftrightarrow (p_y,p_y^\pr)$, symmetries
that will be used in later simplifications of the integrand of $\hard$.

 $\hard$  is now:
\bea
\hard&=& \sum_{s_e,s_{{\overline e}}} \int d^3 p d^3 k d^3 p^\pr d^3 k^\pr
\langle \phi_{1,0,0} | {\bf k} \rangle V^{^{(0)}} ({\bf k},s_3,s_4; {\bf p},s_e,s_{{\overline e}})
\nonumber\\
&&~~~~\times\left(\frac{G_I+G_{II}+G_{III}}{4 m}\right) V^{^{(0)}} ({\bf p}^\pr,s_e,s_{{\overline e}}; 
{\bf k}^\pr,s_1,s_2)
\langle {\bf k}^\pr | \phi_{1,0,0} \rangle \\
&\equiv&\hard(I)+\hard(II)+\hard(III)\;\;~~{\rm respectively}
\;.
\eea
Now we rewrite this in terms of hyperspherical harmonics and perform the
integrations analytically.  The variables are defined as:
 \bea
 &\bullet& \overbrace{[u\equiv(u_o,{\bf u})]}^{\Omega_p}\leftrightarrow[e_1,{\bf p}]\;,\;
 \overbrace{[u^\pr\equiv(u_o^\pr,{\bf u}^\pr)]}^{\Omega_{p^\pr}}\leftrightarrow[e_1,{\bf p}^\pr]
 \label{eq:333}\\
  &\bullet& \underbrace{[v\equiv(v_o,{\bf v})]}_{\Omega_k}\leftrightarrow[e_1,{\bf k}]\;,\;
 \underbrace{[v^\pr\equiv(v_o^\pr,{\bf v}^\pr)]}_{\Omega_{k^\pr}}\leftrightarrow[e_1,{\bf k}^\pr]
 \;.
 \label{eq:444}
 \eea
 See the previous appendix on hyperspherical harmonics for a summary of the
 mathematical relations that we use. The symbols appearing in Eqs.~(\ref{eq:333}) and (\ref{eq:444})
 are explained there.
 Note that we use $e_1$ in these variable definitions, a choice that is completely general
 and turns out to be useful because we are taking expectation values of $n=1$ states in this work.
 The relations  we use are:
\bea
&\bullet&\langle{\bf k}^\pr|\phi_{1,0,0}\rangle=\frac{4 e_1^{\frac{5}{2}}}
{(e_1^2+{{\bf k}^\pr}^2)^2} \frac{1}{\sqrt{2 \pi^2}}\\
&\bullet&\frac{1}{(e_1^2+{{\bf k}^\pr}^2)^2} = \frac{(1+v_o^\pr)^2}{4 e_1^4}\\
&\bullet& d^3 k^\pr=\frac{(e_1^2+{{\bf k}^\pr}^2)^3}{8 e_1^3} d \Omega_{k^\pr}=
\frac{e_1^3}{(1+v_o^\pr)^3} d \Omega_{k^\pr}
\;.
\eea
Given these,  $\hard$ becomes:
\bea
\hard&=& -\frac{m^3 \alpha^5}{32 \pi^2}\int 
\frac{d\Omega_p d\Omega_{p^\pr} d\Omega_k d\Omega_{k^\pr} }
{(1+u_o)(1+u_o^\pr)^2} 
 \left[(\tilde{E}-T)
(G_I+G_{II}+G_{III}) \right]\nonumber\\
&&~~~~~\times{\cal S}\sum_{\mu \mu^\pr} \frac{1}{n n^\pr}Y_\mu(\Omega_p) Y_{\mu^\pr}(\Omega_{p^\pr}) 
Y_\mu^\ast(\Omega_k) Y_{\mu^\pr}^\ast(\Omega_{k^\pr})
\;,
\eea
where
\bea
{\cal S}&\equiv&\sum_{s_e s_{{\overline e}}} v^{(0)}({\bf k},s_3,s_4; {\bf p},s_e,s_{{\overline e}})
v^{(0)}({\bf p}^\pr,s_e,s_{{\overline e}}; 
{\bf k}^\pr,s_1,s_2)
\;.
\eea
Recall Eq.~(\ref{eq:132}) for the definition of $v^{(0)}$.
Using the symmetries of the integrand, the sum over spins
$s_e$ and $s_{{\overline e}}$ can be performed and a  simplification
is seen to arise. The spin completely factors out of the momenta integrations. In other words, 
 we have:
\bea
{\cal S}&=& \frac{1}{6} (3 g_1+ g_2) ({\bf p} \cdot {\bf p}^\pr+{\bf k} \cdot {\bf k}^\pr
-2 {\bf p} \cdot {\bf k}^\pr)
\;,
\label{eq:spring}
\eea 
where
\bea
g_1&\equiv& s_1 s_3+s_2 s_4~,
\label{eq:g1}\\
g_2&\equiv& 1+s_1 s_2-s_2 s_3-s_1 s_4+s_3 s_4+s_1 s_2 s_3 s_4
\;.\label{eq:g2}
\eea
Recall that $s_i = \pm 1$, $(i=1,2,3,4)$; i.e., the `$\frac{1}{2}$' has been factored out of these 
spins.\footnote{
In order to get these simple forms for $g_1$ and $g_2$ and to see this spin/momentum decoupling
it was useful to note the following simple relation: 
$\del_{s s^\pr} = \frac{1}{2} s (s+s^\pr)$ 
 (true because $s^2=1$).} So, in other 
words, instead of having to do sixteen twelve dimensional integrals because the spin and
momenta are coupled together, we just have to do one 
twelve dimensional integral that is independent of spin and then diagonalize the result in the 
$4 \times 4$ dimensional spin space with the spin dependence given by Eq.~(\ref{eq:spring}).

We define the following integral:
\bea
\chi&\equiv&\frac{m \alpha}{8 \pi^2} \xi
\;,
\eea
where
\bea
\xi&\equiv&\int 
\frac{d\Omega_p d\Omega_{p^\pr} d\Omega_k d\Omega_{k^\pr} }
{(1+u_o)(1+u_o^\pr)^2} 
 \left[(\tilde{E}-T)
(G_I+G_{II}+G_{III}) \right]\nonumber\\
&&~~~~~\times(\underbrace{{\bf p} \cdot {\bf p}^\pr}_{a}+\underbrace{{\bf k} \cdot {\bf k}^\pr}_{b}
\underbrace{-2 {\bf p} \cdot {\bf k}^\pr}_{c})\sum_{\mu \mu^\pr}
 \frac{1}{n n^\pr}Y_\mu(\Omega_p) Y_{\mu^\pr}(\Omega_{p^\pr}) 
Y_\mu^\ast(\Omega_k) Y_{\mu^\pr}^\ast(\Omega_{k^\pr})
\;,
\eea
and
\bea
\hard &=& - \frac{m^2 \alpha^4}{24} (3 g_1+ g_2) \chi
\;.
\label{eq:yoyo}
\eea
For the quantities 
$\xi$, $\chi$ and $\hard$, the  labels $I$, $II$ and $III$ imply
the respective terms with 
$G_I$, $G_{II}$ and  $G_{III}$ above  (see Eq.~(\ref{eq:G})). Also,  the
 terms $a$, $b$ and $c$ above  correspond to the respective superscripts in what follows.
This integration will now be performed analytically. 

First the three $G_I$ pieces.  The mathematical relations used here are:
\bea
\del^3(p-p^\pr)&=&\frac{8 e_1^3}{(e_1^2+{\bf p}^2)^3}
 \del(\Omega_p-\Omega_{p^\pr})=\frac{(1+u_o)^3}{e_1^3} \del(\p-\ppr)~, \\
 {\bf p}^2&=&\frac{e_1^2}{1+u_o} (1-u_o)~,\\
 {\bf k} \cdot {\bf k}^\pr &\longrightarrow& 3 k_z k_{z}^\pr~,\\
{\bf p} \cdot {\bf k}^\pr &\longrightarrow& 3 p_z k_{z}^\pr
\;.
\eea
Note that these last two relations are possible due to the rotational symmetry of the integrand.  Then
we expand these z-components of momenta upon the hyperspherical harmonic basis using the following
simple relation (e.g.  the $p_z$ case):
\bea
p_z&=&\frac{e_1}{1+u_o}\left( \frac{\pi i}{\sqrt{2}} Y_{2,1,0} (\p)\right)
\;.\label{eq:252}
\eea
Now we recall the hyperspherical harmonics (see the appendix on hyperspherical harmonics for details)
that we will be using and their orthogonality and phase relationships:
\bea
Y_\mu(\Omega)&\equiv&Y_{n,l,m}(\Omega)\equiv f_{n,l}(\omega) Y_{l,m}(\theta,\phi)~,\\
Y_{n,l,m}&=& (-1)^{l+m} Y_{n,l,-m}^\ast\;,\;Y_{l,m}=(-1)^m Y_{l,-m}^\ast\;,\;f_{n,l}=
(-1)^lf_{n,l}^\ast ~,\\
d\Omega^{^{(4)}}&\equiv& d\Omega \equiv  d\Omega^{^{(3)}} d\omega \sin^2\omega ~,\\
\int d\Omega Y_\mu^\ast Y_{\mu^\pr}&=&\del_{\mu \mu^\pr}\;,\;\int d\omega\sin^2\omega f_{n,l}^\ast
 f_{n^\pr,l}=\del_{n n^\pr}\;,\;\int d\Omega^{^{(3)}} Y_{l,m}^\ast Y_{l^\pr,m^\pr}=
 \del_{l l^\pr} \del_{m m^\pr}
\;.
\eea
After straight-forward application of these relations we obtain:
\bea
&\bullet&\xi_I^a=\frac{4 \pi}{e_1} \int_0^\pi d\omega \sin^2\omega 
\frac{(1-\cos\omega)}{(1+\cos\omega)}=\frac{6 \pi^2}{e_1}
\\&\bullet&\xi_I^b=\frac{3 \pi^2}{2 e_1}\sum_{n=2}^\infty \frac{1}{n^2}
\pig
\\&\bullet&\xi_I^c=-\frac{3 \pi^2}{ e_1}\sum_{n=2}^\infty \frac{1}{n}
\pig
\;.
\eea

For the $G_{II}$ terms, we
 use the following  relations:
\bea
\frac{1}{\tilde{E}-T^\pr}&=&\frac{1}{-\frac{e_1^2}{m}-\frac{{{\bf p}^\pr}^2}{m}}=
-\frac{m}{2 e_1^2} (1+u_o^\pr)~,\\
\frac{1}{\pp}&=&\frac{(1+u_o)(1+u_o^\pr)}{e_1^2}\sum_\mu\frac{2 \pi^2}{n}Y_\mu(\p) Y_\mu^\ast (\ppr)
\;.
\eea
These give:
\bea
(\tilde{E}-T) G_{II}&=&\frac{\alpha m}{2 e_1^4} (1+u_o^\pr)^2(1+u_o)\sum_\mu
 \frac{1}{n} Y_\mu(\p) Y_\mu^\ast(\ppr)
\;.
\eea
We use the rotational symmetry of the integrand and expand the integrand on the hyperspherical harmonic
basis as was done for the three $G_I$ terms. Then we have:
\bea
&\bullet&
\xi_{II}^a = \frac{3 \pi^2}{2 e_1}\sum_{n=2}^\infty\frac{1}{n}\pig
\\&\bullet&
\xi_{II}^b = \frac{3 \pi^2}{2 e_1}\sum_{n=2}^\infty\frac{1}{n^3}\pig
\\&\bullet&
\xi_{II}^c = -\frac{3 \pi^2}{ e_1}\sum_{n=2}^\infty\frac{1}{n^2}\pig
\;.
\eea

 We  use the same  relations for the $G_{III}$ terms and for the three 
$G_{II}$ terms, and we  use the rotational symmetry of the integrand to rewrite the appropriate pieces
of the integrand in terms
of $Y_{2,1,0}$ as we did for the $G_I$ and $G_{II}$ terms. 
However,  we need to discuss one additional relation that allows the 
remaining $\hard(III)$
calculation to
be done analytically. In Schwinger's 1964 paper~\cite{schwinger}
he gives the following formula:
\bea
\frac{1}{2 \pi^2} \frac{1}{(1-\rho)^2+\rho(u-u^\pr)^2}&=&
\sum_{n=1}^\infty \rho^{n-1} 
\frac{1}{ n} \sum_{l,m} Y_{n,l,m}(\Omega) Y_{n,l,m}^\ast (\Omega^\pr)
\;,\label{eq:billy1}
\eea
where $u$ and $u^\pr$ are of unit length and $0 < \rho 
< 1$.\footnote{This is easily derivable from a more general standard
formula that 
Schwinger gives,
\beaa
\frac{1}{4 \pi^2}\frac{1}{(u-u^\pr)^2}
=\sum_{n=1}^\infty \frac{\rho_{<}^{n-1}}{\rho_{>}^{n+1}} 
\frac{1}{2 n} \sum_{l,m} Y_{n,l,m}(\Omega) Y_{n,l,m}^\ast (\Omega^\pr)
\;.\eeaa}
Inside the brackets in $G_{III}$ we have:
\bea
&&\!\!\!\!\!\!\!\!\!\!\left[\int_{0}^{1}\frac{ d \rho }{\rho} \left(
\frac{1}{4 e_1^2 \rho \pp+C (1-\rho)^2}-
\frac{1}{C}
\right)
\right] = \left[\int_{0}^{1}\frac{ d \rho }{\rho}\frac{1}{C}\left(\frac{1}{{(1-\rho)^2+
\rho(u-u^\pr)^2}}-1\right)\right]
\;.\label{eq:247}
\eea
Recall $C\equiv (e_1^2+{\bf p}^2)(e_1^2+{{\bf p}^\pr}^2)$.
Also recall that we are using the coordinate change of Eqs.~(\ref{eq:333}) and (\ref{eq:444}).
Eq.~(\ref{eq:1492}) with $e_n = e_1$ then applies and was used.
In  Eq.~(\ref{eq:247}), $0 < \rho < 1$ and $u$ and $u^\pr$ are of unit length, thus 
 Schwinger's equation
can be used and we have:
\bea
(\tilde{E}-T) G_{III}&=&\frac{\alpha m (1+u_o)(1+u_o^\pr)^2}{2 e_1^4}
\int_0^1 d\rho\sum_{\mu \neq (1,0,0)} \frac{\rho^{n-2}}{n}Y_\mu(\p) Y_\mu^\ast (\ppr)\;.
\label{eq:billy2}
\eea
Now, since $n \geq 2$ in this sum we can do the integral over $\rho$:
\bea
\int_0^1 d\rho \rho^{n-2}=\left.\frac{\rho^{n-1}}{n-1}\right|^1_0 = \frac{1}{n-1}
\;,
\eea
 and we obtain:
\bea
(\tilde{E}-T) G_{III}&=&\frac{\alpha m (1+u_o)(1+u_o^\pr)^2}{2 e_1^4}
 \sum_{\mu \neq (1,0,0)} \frac{1}{n (n-1)}Y_\mu(\p) Y_\mu^\ast (\ppr)\;.
\eea
For terms in $\xi$ which contain $G_{III}$, one obtains:
\bea
&\bullet&
\xi_{III}^a = \frac{3 \pi^2}{2 e_1}\sum_{n=2}^\infty\frac{1}{n(n-1)}\pig
\\&\bullet&
\xi_{III}^b = \frac{3 \pi^2}{2 e_1}\sum_{n=2}^\infty\frac{1}{n^3 (n-1)}\pig
\\&\bullet&
\xi_{III}^c = -\frac{3 \pi^2}{ e_1}\sum_{n=2}^\infty\frac{1}{n^2 (n-1)}\pig
\;.
\eea

Now recall  $\chi\equiv\frac{m \alpha}{8 \pi^2} \xi$, and
 also notice  that all the summands are the same, thus putting it all together we have:
 \bea
 \chi&=&\frac{3}{2}+\frac{3}{8}\sum_{n=2}^\infty
 \left(
 \frac{1}{n}+\frac{1}{n (n-1)}+\frac{1}{n^2}
 +\frac{1}{n^3}+\frac{1}{n^3 (n-1)}\right.\nonumber\\
 &&~~~~~- \left.\frac{2}{n}
 -\frac{2}{n^2}-\frac{2}{n^2 (n-1)}
\right) \pig\\
&=&\frac{3}{2}-\frac{3}{8}\sum_{n=2}^\infty
 \left(
 \frac{1}{n}
\right) \pig
 \;.
 \label{eq:kkk}
 \eea
 The remaining sum can be done analytically.
 To see this, first define two integrals:
 \bea
 I_1&\equiv&\int \frac{d\Omega_p}{1+u_o} \frac{d\Omega_{p^\pr}}{1+u_o^\pr}=
 (4 \pi)^2\left(\int_0^\pi \frac{d\omega \sin^2 \omega}{1+\cos\omega}\right)^2=
 16 \pi^4~,\label{eq:eeooe}\\
 I_2&\equiv&\int \frac{d\Omega_p}{1+u_o} \frac{d\Omega_{p^\pr}}{1+u_o^\pr} 
 \frac{(\kap-\kap^\pr)^2}{\pp}=
 \frac{2}{3}I_1
 \;.\label{eq:sss}
 \eea
 The last equality followed from rotational symmetry of the integrand.
 We can also calculate $I_2$ a hard way which gives:\footnote{
 We use Eq.~(\ref{eq:252}), $p_x=\frac{e_1}{1+u_o} \frac{\pi i}{2}
 \left(Y_{2,1,-1}(\Omega_p)-Y_{2,1,1}(\Omega_p)\right)$ and 
 $p_y=-\frac{e_1}{1+u_o} \frac{\pi}{2}\left(Y_{2,1,-1}(\Omega_p)+Y_{2,1,1}(\Omega_p)\right) $
 .}
 \bea
 I_2&=&16 \pi^4-4\pi^4 \sum_{n=2}^\infty\frac{1}{n} \pig=\frac
 {32 \pi^4}{3}
 \;.
 \eea
 This last equality followed from Eqs.~(\ref{eq:eeooe})~and~(\ref{eq:sss}) (the easy 
 way to calculate $I_2$).
 Thus, we have:
 \bea
&&\sum_{n=2}^\infty\frac{1}{n} \pig=\frac{4}{3}
 \;.
 \eea
 
  Combining this result with Eq.~(\ref{eq:kkk}) gives:
\bea
 \chi&=&\frac{3}{2}-\frac{3}{8}\left( \frac{4}{3}\right)=1
 \;.
 \eea
 Thus, recalling Eq.~(\ref{eq:yoyo}), we have:
 \bea
 \hard &=& - \frac{m^2 \alpha^4}{24} (3 g_1+ g_2) 
 \;,
 \label{eq:whatthehell}
 \eea
 where $g_1$ and $g_2$ are given in Eqs.~(\ref{eq:g1}) and (\ref{eq:g2}) respectively.
%%%%%%%%%%%%%%%%%%%%%%%%%%%%%%%%%%%%%%%%%%%%%%%%%%%%%%%%%%%%%%%%%%%%%%%%%%
%%%%%%%%%%%%%%%%%%%%%%%%%%%%%%%%%%%%%%%%%%%%%%%%%%%%%%%%%%%%%%%%%%%%%%%%%%%%%%%
\vskip.2in
\noindent
\centerline{{\bf  APPENDIX D:} {\boldmath $M_{_{N}}^2$} {\bf -VS-} {\boldmath $B_{_{N}}$} }
\vskip.2in

In this appendix we will invert the equation
\bea
M_{_{N}}^2&\equiv&(2 m+B_{_{N}})^2
\;,
\eea
and obtain the $\alpha$-expansion for
 the binding energy, $B_{_{N}}$. In this work we set up a procedure to calculate $M_{_{N}}^2$. This gave
 \bea
 M_{_{N}}^2&=&{\cal M}_N^2+b_4 m^2 \alpha^4+{\cal O}\left(\alpha^5\right)\;,
 \label{eq:eq}
 \eea
 with
 \bea 
 {\cal M}_N^2&\equiv&4 m^2+4 m {\cal B}_{_{N}}\;.
 \eea
 For the lowest order spectrum of $H_{_{R}}$ we obtained
 \bea
 {\cal B}_{_{N}}&=& -\frac{1}{4} \frac{m \alpha^2}{n^2}\;.
 \eea
Taking a square root of $M_{_{N}}^2$ gives
 \bea
 B_{_{N}}&=& {\cal B}_{_{N}}
 +\frac{m \alpha^4}{2} \left( \frac{b_4}{2}-\frac{1}{32 n^4}     \right)
+{\cal O}\left(\alpha^5\right)
 \;.\label{eq:eee}
 \eea
 Recall that $\frac{m \alpha^4}{2}=\alpha^2 Ryd$. Now, in this work,
  Eqs.~(\ref{eq:1776})-(\ref{eq:1493}) 
 are the results of our calculation of the spin splittings of $M_{_{N}}^2$ in the ground state of
 positronium. These 
 were derived in the form of Eq.~(\ref{eq:eq}) with result
 \bea
 b_4(triplet)-b_4(singlet)&=&\frac{7}{3}
 \;.
 \eea
 Given Eq.~(\ref{eq:eee}) this implies
 \bea
 B_{_{triplet}}-B_{_{singlet}}&=&\frac{7}{6} \alpha^2 Ryd +{\cal O}\left(\alpha^5\right)
 \;.
 \eea
 This we recognize as the well known result for the positronium system.  A final note is
 that if the physical values of the fine structure constant and Rydberg energy 
 ($\frac{1}{137.0}$ and $13.60 eV$ respectively)
 are applied to this previous formula, the result agrees with experiment to one-half of a
  percent~\cite{IZ2}.
%%%%%%%%%%%%%%%%%%%%%%%%%%%%%%%%%%%%%%%%%%%%%%%%%%%%%%%%%%%%%%%%%%%%%%%%%%%%%%%%%%%%%%%%%%%%%%%%%%%%%%%%%%%%%%%%%%%
%%%%%%%%%%%%
%%%%%%%%%%%%
%%%%%%%%%%%%                here's the references
%%%%%%%%%%%%
%%%%%%%%%%%%%%%%%%%%%%%%%%%%%%%%%%%%%%%%%%%%%%%%%%%%%%%%%%%%%%%%%%%%%%%%%%%%%%

\underline{Figure captions}

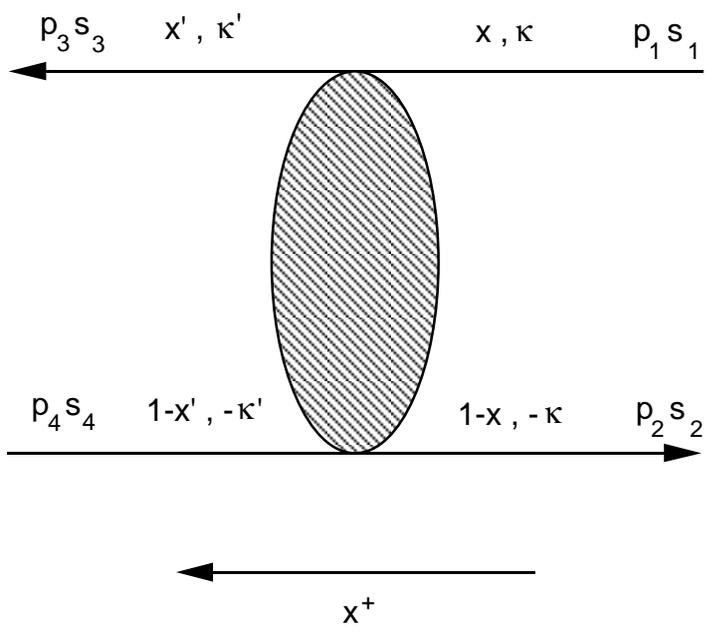
Figure 1:  This illustrates the spin and momenta label conventions used in this paper.

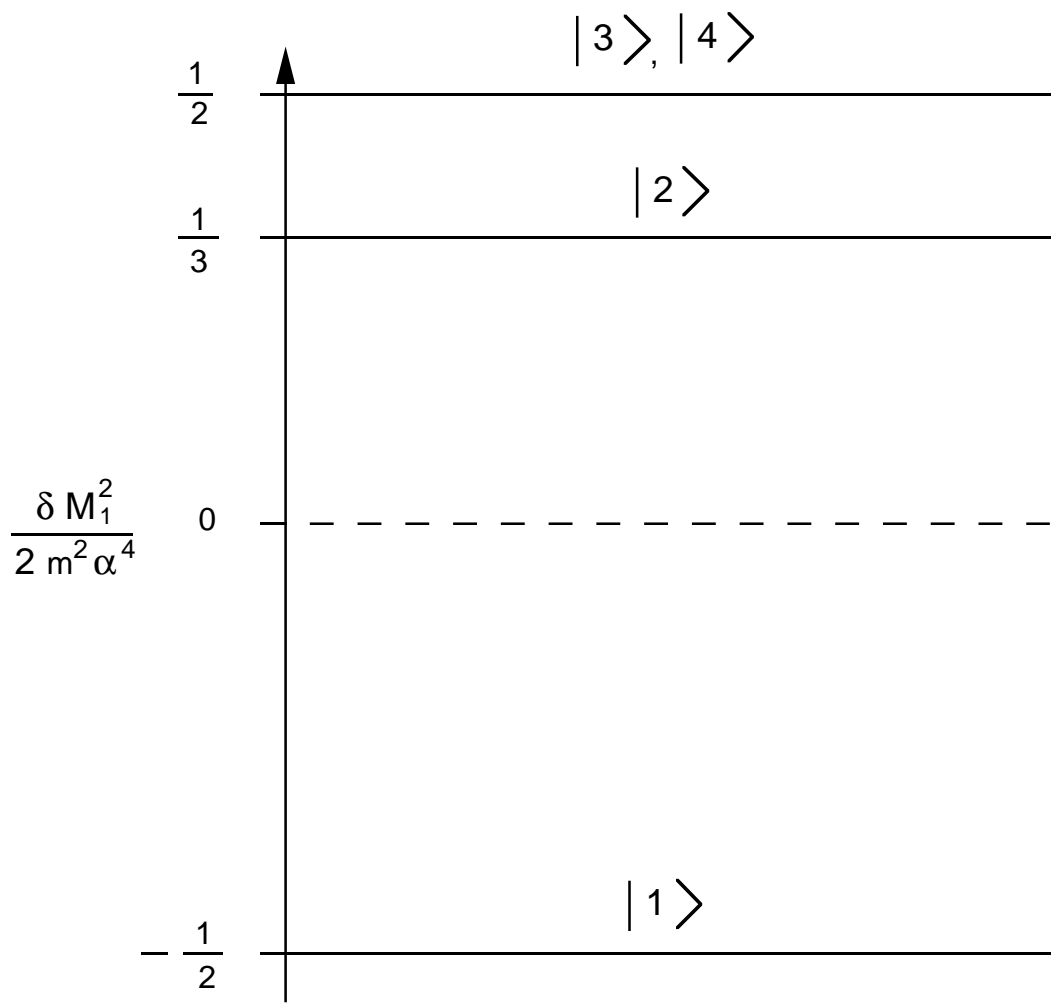
Figure 2: $\del M_{_{1}}^2$ is the part of the ground state mass spin splittings 
from Eq.~(\ref{eq:178}). $m$ is the electron mass and $\alpha$ is the fine structure constant.
The state labels 1, 2, 3 and 4 are explained in Eq.~(\ref{eq:187}). The two upper most
levels should coincide in a rotationally invariant theory.

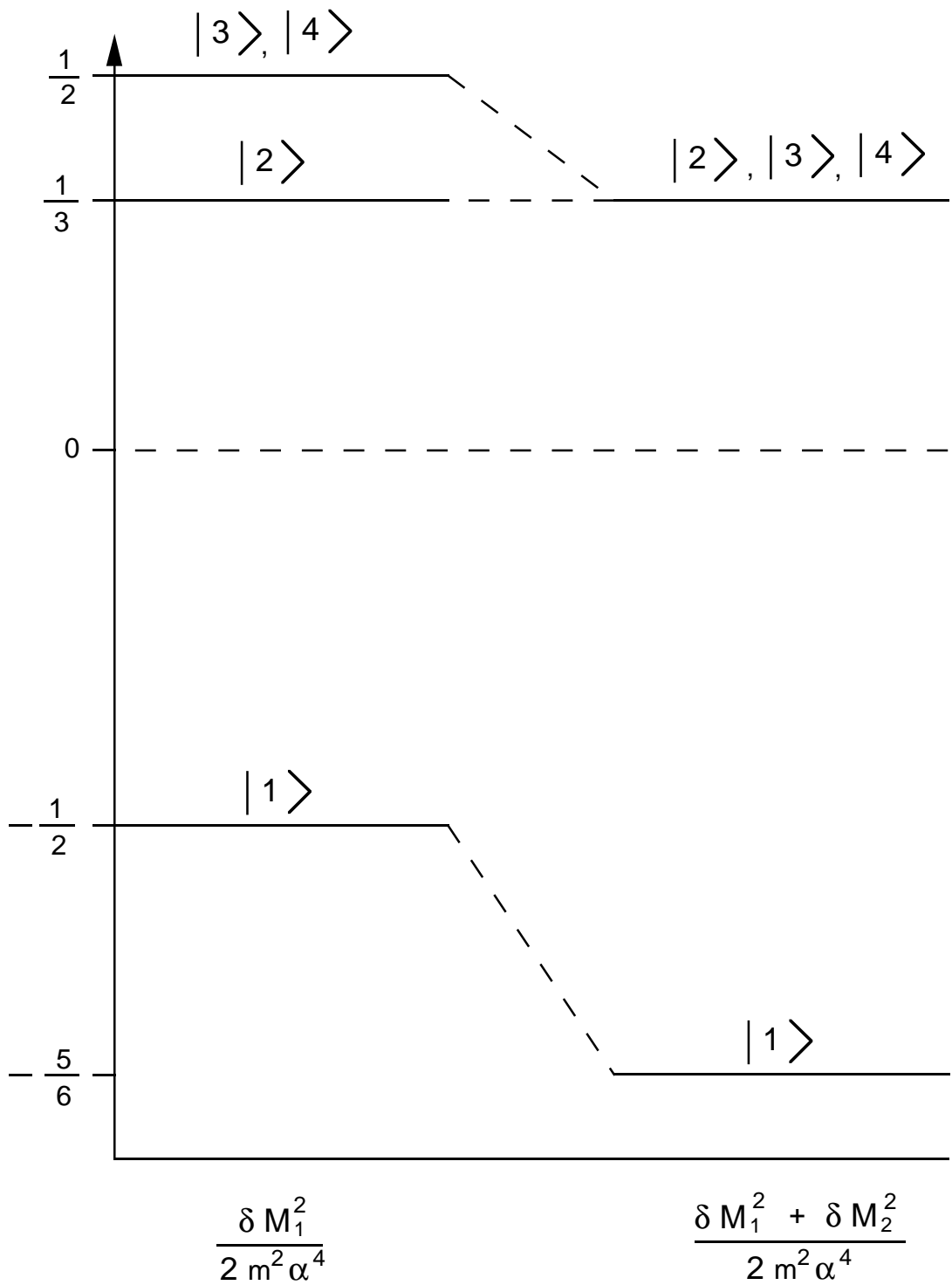
Figure 3: The combined ground state mass spin splitting in positronium 
to order $\alpha^4$ is illustrated 
using the same notation as in Figure 2. $\del M_{_{2}}^2$ is given by Eq.~(\ref{eq:happyman}) and is 
calculated in Appendix~C. The final combined result  (on the right) corresponds
to a rotationally invariant theory.

\end{document}